# Malware-Resistant Data Protection in Hyper-connected Networks: A survey

Jannatul Ferdous*, Rafiqul Islam, Maumita Bhattacharya and Md Zahidul Islam

*School of Computing, Mathematics, and Engineering, Charles Sturt University, NSW, Australia*

**Cyber Security Cooperative Research Centre, Australia*

**Abstract**

Data protection is the process of securing sensitive information from being corrupted, compromised, or lost. A hyper-connected network, on the other hand, is a computer networking trend in which communication occurs over a network. However, what about malware? Malware is malicious software meant to penetrate private data, threaten a computer system, or gain unauthorized network access without the user's consent. Due to the increasing applications of computers and dependency on electronically saved private data, malware attacks on sensitive information have become a dangerous issue for individuals and organizations across the world. Hence, malware defense is critical for keeping our computer systems and data protected. Many recent survey articles have focused on either malware detection systems or single attacking strategies variously. To the best of our knowledge, no survey paper demonstrates malware attack patterns and defense strategies combinedly. Through this survey, this paper aims to address this issue by merging diverse malicious attack patterns and machine learning (ML) based detection models for modern and sophisticated malware. In doing so, we focus on the taxonomy of malware attack patterns based on four fundamental dimensions: the primary goal of the attack, method of attack, targeted exposure and execution process, and types of malware that perform each attack. Detailed information on malware analysis approaches is also investigated. In addition, existing malware detection techniques employing feature extraction and ML algorithms are discussed extensively. Finally, it discusses research difficulties and unsolved problems, including future research directions.

## 1. Introduction

In this digital world, preserving the security of sensitive data can be challenging for internet users due to the threat of unauthorized computer system access by malware attacks. Malicious programs or malware are any undesirable programs or files that are secretly injected into other hosts and are committed to destroying the operating system or network and leading to the exfiltration of private data and other harmful effects. Cyber attackers utilize various malicious programs, such as ransomware, spyware, adware, rootkit, worm, horse, botnets, trojans, and viruses at different times to convert the contents of a computer's file system without the victim's consent. Malware attacks can compromise a computer system using various techniques, including spreading from compromised systems, deceiving users into downloading malicious files and attracting users to access malicious websites. Malware targets include end-user computers, network equipment (such as routers and switches), servers, and logic controllers. Today's modern internet is afflicted by the growth and intelligence of an ever-increasing quantity of malware [1].

The digital revolution has grown increasingly important in our daily lives due to its better efficiency, rapid communication, and incomparable simplicity. People can share information and transfer money with just a click. Unfortunately, dealing with malware attacks remains a challenge for the netizen, even with advances in technology and cybersecurity, because the threat actors are always trying to discover new ideas for making cash or rising trades by thieving private data, bank accounts, and credit statements from many public and private organizations. This increases the considerable safety risk of data privacy for the user. According to Cybersecurity Statistics, 653% of malicious activity was claimed in July 2020 alone, compared to the same month in 2019 [2]. Another report by the US FBI has shown that in 2021, the number of malicious threats climbed by around 300% due to the growing number of internet users, particularly during the COVID -19 pandemic [3]. Therefore, this creates a significant potential for vast financial losses globally. Cybersecurity Ventures estimated that in 2021 the worldwide fiscal loss due to cybercrime was about $6 trillion. Moreover, to smooth data accessibility and the distribution of computer resources, many organizations, governments, and enterprises usually gather and keep sensitive data in a host machine. If malware attacks infect an organization's host computers, they may share sensitive data and many things to blend in during the execution processes. Hence, identifying harmful run-time attempts and other attacks is critical to protecting sensitive data while sharing in hyper-connected networks.

Researchers have presented many solutions to control and mitigate malware attacks using machine learning or other techniques to defend the security and confidentiality of private information. However, these processes can be challenging because today's malware threats are sneakier, more detrimental, and spread to other hosts silently. Hence, it is more difficult to spot and eliminate than earlier generations of malware [4]. Furthermore, cyber attackers use emerging sophisticated and different obfuscation techniques nonstop to design advanced malware variants, such as polymorphism, metamorphism, oligotrophic, etc., to fool security experts [5]. They develop sophisticated programming in such a method that disturbs the operation process. In addition, while executing in a controlled environment, advanced malware may detect and bypass anti-malware tools and hide destructive features. Hence, malware developers and anti-malware detection systems are locked in a never-ending arms race [6]. Furthermore, we regularly observe a rising number of zero-day attacks due to the dynamic nature of malware attacks. Zero-day attacks are attacks that have never been seen before. All the above factors are making the malware detection process more difficult. However, although many surveys in malware research have already been done [5],[7]-[16], these are either outdated or limited in scope. More specially, no comprehensive survey identifies both diverse malware attacks and their detection methods, which is the central issue of this review paper.

This paper has conducted comprehensive and in-depth surveys of the existing papers focused on an overview of malware, the taxonomy of malware attack patterns, and three malware detection techniques, namely static-based, dynamic-based, as well as hybrid-based. Various feature extraction procedures and classification algorithms are discussed and reviewed to find an effective and robust method for classifying and identifying malicious programs. This survey is important because it is a concise framework that encompasses all areas of malware and gives vast relevant information.

### *1.1 Contributions*

Malware is considered one of the leading threats among internet users today, particularly during this COVID−19 pandemic. Therefore, it is crucial to analyze and detect malware to defend against malicious attacks and stop their detrimental acts. The key contributions of this paper are listed below:

- A taxonomy of malware attack patterns is developed that divides them into categories such as Polymorphic & Metamorphic attacks, Ransomware attacks, Fileless malware attacks, Advanced Persistence Threats (APT), and Zero-day attacks based on the targeted exposure, the attack techniques, execution process, and the forms of malware used to carry out the attack.
- A comparative analysis has been presented using static, dynamic, and hybrid analysis methods and their tools

and techniques to capture malware behavior.
- Various types of static and dynamic feature extraction methods are discussed and reviewed.
- A comparative evaluation of various machine-learning methods used in malware classification has been presented.
- We have explored various research difficulties and challenges in this study, all of which are important in the performance of malware classifiers.

### *1.2 Scope*

Selecting an appropriate and efficient malware detection approach is an open challenge. Hence, new approaches and experimental studies are very crucial in the anti-malware community. As a result, this survey paper will serve as a benchmark for developing a prototype for detecting cyber-attacks and responding effectively. In addition, this paper is designed to help cybersecurity experts who are keen on using ML techniques to automate the malware analysis process.

### *1.3 Organization*

Before surveying this study, Section 2 illustrates an overall comparison of our survey with previous surveys, followed by Section 3 on the basics of malware. Sections 4 and 5 provide a broad review of existing research, including malware attack taxonomy and malware detection procedures, respectively. Section 6 summarizes and compares the papers on malware detection methods based on the surveyed input features and ML classifiers. Section 7 focuses on the research issues and challenges, which effectively encapsulate the strong and weak points of the findings or an assessment of the survey's excellence. It also discusses some limitations that remain unsolved. Finally, Section 8 brings this paperto a conclusion, including future research directions. Figure 1 presents the complete structure of this paper.

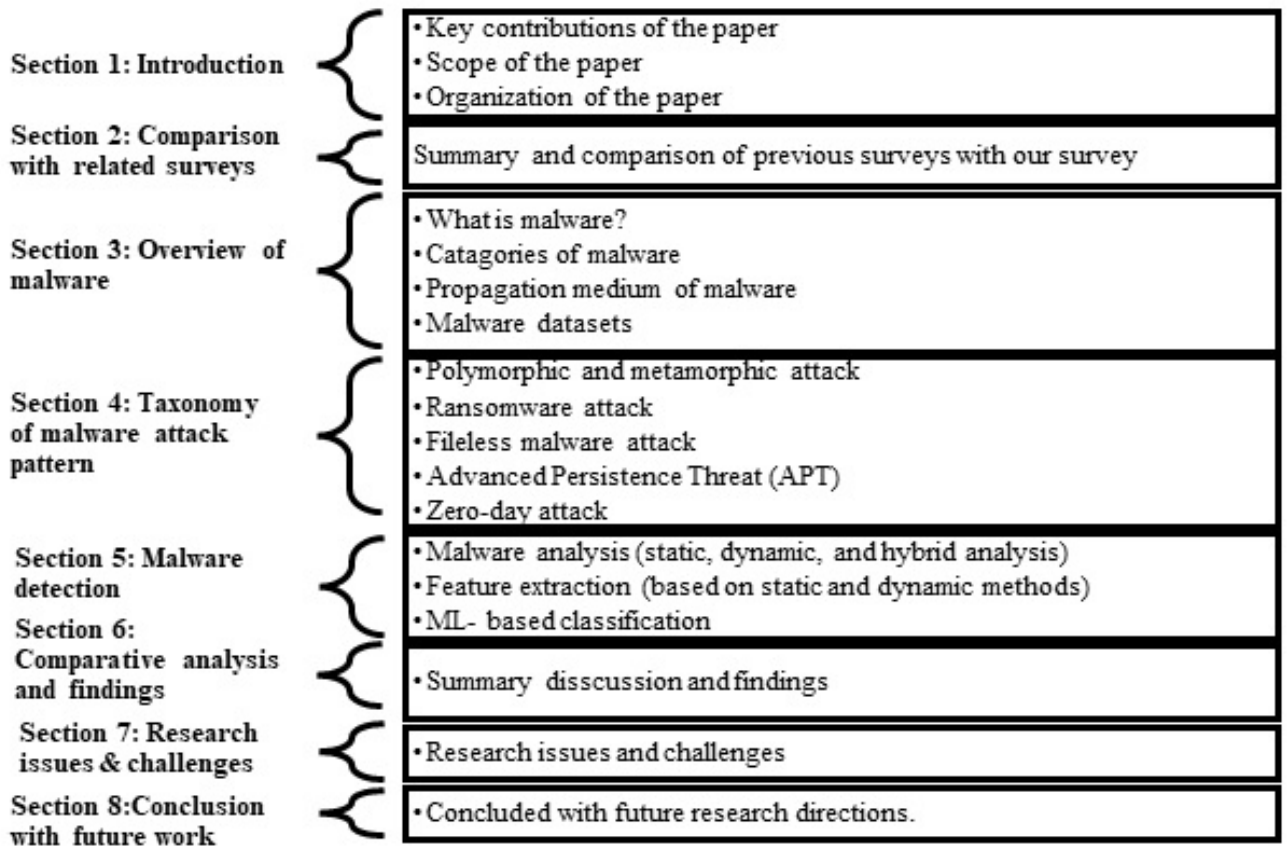

Figure 1: Complete outline of the paper

## 2. Comparison with related surveys

This section summarizes the reviewed literature on malware analysis from 2011 to 2021 and analyzes the shortcomings that we aim to address in our work shown in Table 1. It will help researchers to construct a baseline for developing a method to counter such attacks. There has already been much research done on the use of machine-learning approaches for malware classification. Shabtai et al. (2009) produced the first study on this issue. Souri and Hosseini (2018) [7] offered an arrangement of machine learning-based malware-hunting techniques. Their study varies from our proposed paper in that they do not explore which features are taken seriously. Bazrafshan et al. (2013) [8] and Basu (2016) [9] concentrated on malware detection using machine learning and other techniques; however, they used a restricted number of feature types, whereas this paper suggests a higher number of feature types, highlighting the broader scope of our research. A complete investigation of the development and present scenario for the detection of malicious code based on ML techniques was proposed by Singh and Singh (2021) [10], Sihwail et al. (2018) [11], Aslan and Samet (2020) [5], Abijah Roseline and Geetha (2021) [12], and Abusitta et al. (2021) [13]. However, they did not mention anything about the attack pattern. Other studies, for example, Gandotra et al. (2014) [14], Singla et al. (2015) [15], Yu et al. (2018) [16], Choudhary and Sharma (2020) [17], and Caviglione et al. (2021) [6] examined various machine learning methods for malware classification based on static and dynamic detection techniques, whereas our survey focused on hybrid malware detection including static and dynamic approaches. Moreover, some other surveys focused on only various dynamic analysis tools and techniques and malware classification taxonomy like Egele et al. (2012) [18], Or-Meir et al. (2019) [19], and Talukder and Talukder (2020) [20].

Table 1: Summary and comparison of previous surveys with our study (myth: √ = Complete information covered, ≈ = Partial information covered, × = No information covered).

| Authors & year | Analysis methods | | Detection methods | | | Malware attack pattern | | | | |
|---|---|---|---|---|---|---|---|---|---|---|
| | Static analysis | Dynamic analysis | Static–based | Dynamic-based | Hybrid-based | Ransomware attack | Polymorphic & metamorphic attack | Fileless malware attack | Advanced Persistent Threat | Zero-day attack |
| Singh and Singh (2021) [10]; Sihwail et al. (2018) [11]; Aslan & Samet (2020) [5]; Abijah Roseline & Geetha (2021) [21]; Abusitta et al. (2021) [13]; Singh & Singh (2021) [10]. | √ | √ | √ | √ | √ | × | × | × | × | × |
| Gandotra et al. (2014) [14]; Singla et al. (2015) [15]; Yu et al. (2018) [16]; Choudhary & Sharma (2020) [17];Caviglione et al. (2021) [6]. | √ | √ | √ | √ | × | × | × | × | × | × |
| Egele et al. (2012) [18]; Or-Meir et al. (2019) [19]; Talukder & Talukder (2020) [20]. | × | √ | × | √ | × | × | × | × | × | × |
| Berrueta et al. (2019) [22]; Harun Oz et al. (2021) [23]; and Moussaileb et al. (2021) [24]. | √ | √ | ≈ | √ | ≈ | √ | × | × | × | × |
| Sharma & K. Sahay (2014) [25]. | √ | √ | √ | √ | √ | × | √ | × | × | × |
| Sibi Chakkaravarthy et al. (2019) [26]. | √ | √ | √ | √ | ≈ | × | × | × | √ | × |
| Kaur & Singh (2014) [27]. | √ | √ | ≈ | × | × | × | × | × | × | √ |
| Sudhakar & Kumar (2020) [28]. | × | √ | × | √ | × | × | × | √ | × | × |
| Our survey | √ | √ | √ | √ | √ | √ | √ | √ | √ | √ |

However, all researchers explained only a single malicious attack pattern. For example, Berrueta et al. (2019) [22], Harun Oz et al. (2021) [23], and Moussaileb et al. (2021) [24] mainly focused on the classification and detection of cryptographic ransomware employing ML algorithms. Sharma and K. Sahay (2014) [25] proposed a detailed survey for detecting and classifying polymorphic and metamorphic malware to protect data from their corresponding attacks. Kaur and Singh (2014) [27] conducted a detailed survey on zero-day attacks. Sibi Chakkaravarthy et al. (2019) [26] concentrated on innovative Advanced Persistent Threat (APT) and the study of Sudhakar and Kumar (2020) [28] examines all fileless malware's behavior. It is apparent from the literature that most recent studies do not provide in-depth research or offer remedies in a particular area. To tackle this problem, this paper presents detailed information on feature extraction, ML algorithms, and malware detection techniques to make our study as simple and informative as possible for the reader. However, most of the survey articles focus solely on the same attack in different ways. For instance, some researchers focus on ransomware attacks, others emphasize zero-day attacks and so on for other attacks.

Our survey addresses this gap by combining and simplifying their work's diverse attack patterns. Thus, our survey is comprehensive and unique as it provides wide-ranging insights into various attack patterns and malware detection systems, as opposed to other surveys.

## 3. Overview of malware

This section provides an overview of malware, including how malware is defined, malware categories, malware spreading mediums, and sources of collecting malware datasets.

### *3.1 What is malware?*

Malware, termed "malicious software" is a type of computer program with the intent of harming or exploiting another software application or compatible devices. Malware began to appear in the 1980s and the very first computer malware, dubbed Brain, was launched in 1986 [29]. A malicious program can be distributed in various formats, including executables, binaries, shellcodes, scripts, and software. In this paper, we use the terms "binary code", "malignant scripts" or "malicious program" to represent malware [19]. Malware attacks can break security holes, penetrate deeper into devices, propagate across networks, and interrupt an organization's main ongoing operations. Malware is the primary cause of most attacks, including significant data breaches that result in widespread fraud and identity theft. Ransomware attacks, which cost thousands of dollars, are also driven by malware. Cybercriminals target individuals, businesses, and even government agencies with malicious attacks [30].

### *3.2 Categories of malware*

Malware can be categorized into several forms depending on its aims and the distribution of information systems. Table 2 depicts a brief description of diverse sorts of malware.

*Classification of malware by malicious behavior*

The classifications and concepts listed above are suitable when describing malware to non-professionals. However, for analysis, it is more crucial to concentrate on malware behavior rather than malware type which is as follows [19].

*Stealing information*-Theft of information is the most typical malicious action, which can involve financial information, personal information, passcodes, or access credentials. According to the CIA triad, data confidentiality is undermined by information theft, which is most associated with malware like trojans, spyware, etc.

*Creating a vulnerability*-Malware can generate new vulnerabilities by disabling anti-virus software, installing spyware, altering usernames, modifying firewall policies, degrading software to an outdated version, and other methods. This activity endangers the system's security and is linked to RATs (Remote Access Trojans) and Bots.

*Denying service*- A denial-of-service attack can be risky when services are frequently visited, as they are now because it reduces service availability. Hackers can refuse service in a variety of ways.

*Executing commands from the C&C*- Malware developers occasionally use the C&C server to send and receive information to and from the victim's computer which allows it to carry out malicious activities. This type of behavior, which is commonly linked with bots, ransomware, and RATs, threatens the system's integrity.

*Deceiving the user*- Fraud can be exploited by bad actors to enter secured systems and/or manipulate data for their benefit. This activity is linked to Trojans, RATs, and scareware, and it put at risk the integrity and confidentiality of the system.

*Stealing computing resources*- Apart from doing the computations required to mine bitcoins, crypto miners generally enable the system to run normally and do not interact with the system's information. This activity puts at risk the system's integrity and availability.

*Spreading*-A common behavior seen in worms and viruses is spreading.

Figure 2 depicts the malicious behavior that each type of malware exhibits. The link between the standard malware taxonomy and the behavioral taxonomy we presented here is depicted in this diagram.

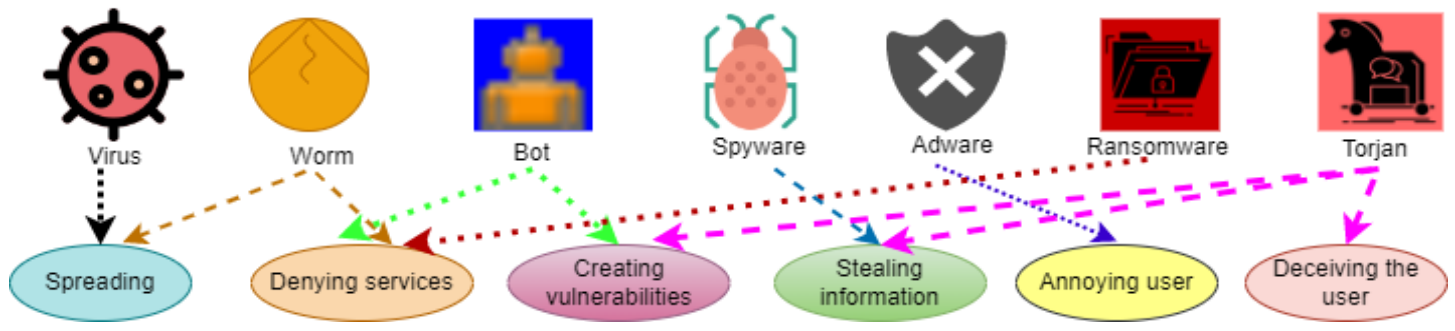

Fig 2: Correlation between malware types with malicious behavior.

Table 2: Classification of malware depending on the purpose and information-sharing system.

| Malware types | Brief description | Examples |
|---|---|---|
| Virus [21] | It carries out harmful operations such as removing or altering system or user data files. | Sector virus, Brain boot, Elk Cloner, etc. |
| Worm [31] | A worm is computer software that copies itself and spreads over networks. | Morris, Blaster, Melissa, Stuxnet, My doom, Sasser, etc. |
| Rootkit [31] | A rootkit is a piece of malware that owns a victim's machine from a distance without being detected by the user. | NTRootkit, SONY BMG, Copy Protection rootkit. |
| Trojan horse [21] | A phishing program that tries to pass itself off as a harmless application, a good thing, or even fun to stay hidden and carry out its nefarious activities. | Trojan-Banker, Trojan-Downloader, Trojan-Dropper, etc. |
| Adware [21] | Adware displays advertising to the user automatically. Adware can inject this advertising into other computer software or web links; in certain situations, it can even replace a prevailing advertisement with a new one. | Plankton, Fireball, Dollar Revenue, Gator, DeskAd, etc. |
| Spyware [31] | Spyware monitors user activity invisibly and without the user's knowledge. | Finisher, internet optimizer, Look2Me, etc. |
| Botnet [32] | Bot malware compromises computer systems to exploit their resources. | Agobot, Mirai, Conficker, Zeus, Waledac, etc. |
| Backdoor [33] | A computer program is created to avoid a computer's security features and implant them on a device, allowing a threat actor to enter the computer. | – |
| Ransomware [31] | The ransomware code locks the users' personal information and asks for money (a ransom) to get it back. | WannaCry, CryptoLocker, Cryptowall, etc. |

### 3.3 *Propagation medium of malware*

Malware propagates throughout host systems due to human actions, both directly and indirectly, and cybercriminals are constantly devising new ways to infiltrate a victim's system to access sensitive data. Some of the most popular sources for malware propagation are as follows:

*Drive-by Download* - Anyone connected to the network and engaging in surfing websites is a possible victim. Users can upload malware-infected-pirated programs that appear genuine [1].

*Network*- Malicious actors sometimes use the network as a source of malicious code to perform the attack operation because malware can propagate throughout any network [34].

*Hacked websites*-Hackers can upload harmful code to a legitimate site if the user is unaware of system vulnerabilities in the web's configuration.

*Backdoors*- Backdoors are means of access to a computer through which malware programs are installed. A well-known backdoor is FinSpy. Once implemented on a system, it allows an attacker to upload and execute malicious scripts on the machine remotely [6].

*Ads*- Advertisements are another tricky source of malicious attacks for cybercriminals. A user can infect the computer with malware by clicking on advertisements on relevant websites.

*Phishing*- Phishing is a popular source of malware to conduct the attack procedure. The attackers present themselves as legitimate entity and urge you to offer private information. Users may become infected with malware by clicking on these kinds of links or movies [30].
*Removable drives*- Flash drives and hard disks are examples of removable drives. These are the most common methods for malware propagating from one machine to another. Removable drives can distribute any malware, including viruses, worms, and ransomware [6].
*Software downloads*- Software downloads can also be a source of malware as users can obtain a wide range of useful software via the internet.

### *3.4 Malware datasets*

To understand malware's tricks and strategies, researchers need to collect malware datasets. One method of gathering samples is by online sites of anti-malware projects like MalShare, Malware DB, VirusShare, etc. Additionally, some specific organizations and research workgroups sometimes try sharing their malware sample data to address the lack of publicly available data sources, for example, Microsoft, Ember, etc. Table 3 presents some open-source links that can be used to gather malware samples.

Table 3: List of some publicly available open sources for collecting malware samples.

| Sources | Description | Dataset weblink |
| --- | --- | --- |
| Microsoft Malware Classification Challenge | It was released by Microsoft and contained more than 20,000 pieces of malware from nine different malware types with a combination of 10,868 Bytes files and 10,868 ASM files [35]. | https://www.kaggle.com/c/Malware-classification |
| EMBER dataset (2018) | Endgame Malware Benchmark for Research has 1 million binary samples, including 900K training data and 200k test data. The dataset is available for developing models that can identify malicious Windows PE files [36]. | https://github.com/endgameinc/ember. |
| VirusShare | The VirusShare dataset is a great publicly available source for researching malware [37]. | VirusShare.com |
| Malshare | The MalShare Project is a collaborative venture that offers public access to malware samples and allows 2000 calls per day. This project benefits everyone by providing free resources and permitting high-volume use [38]. | MalShare |
| SoReL-20M | Sophos and Reversing Labs have compiled a database containing 20 million malicious files as well as 10 million disabled malware applications, in response to the lack of reliable data [39]. | https://github.com/sophos-ai/SOREL-20M |
| SANDBOX | The Cuckoo Sandbox produced malicious public files for computer security researchers, based on a study of Windows API calls [40]. | https://github.com/ocatak/malware_api_class |
| MalwareBazaar | The MalwareBazaar classifies specimens according to the date, data type, signature, and other information [41]. | MalwareBazaar \| Malware sample exchange (abuse.ch) |
| InQuest Malware | This malware database provides a variety of malicious apps and information about their analysis [42]. | InQuest/malware-samples |
| Contagio Malware Dump | Contagio is a collection of the newest malicious files, attacks, discoveries, and assessments [43]. | http://contagiodump.blogspot.com/ |
| theZoo | A project called "The Zoo" was created to make malware detection publicly accessible [44]. | https://github.com/ytisf/theZoo |
| VirusSign | VirusSign provides 100,000 different types of malware every day for researchers [45]. | https://www.virussign.com/downloads.html |

## 4. Taxonomy of attack pattern

An attack pattern is a conceptualization technique that outlines how a specific sort of detected threat is carried out. It helps security experts and designers to understand how their systems can be exploited and how to protect them successfully. The key points should be included in an attack pattern namely-Name and classification of the attack pattern, targeted exposures, or vulnerabilities, attacking method, attacker goal, and consequences, and how malware attack works [46]. Figure 3 shows the taxonomy of malware attack patterns. It is helpful to generate a taxonomy to characterize the huge range of malicious attacks systematically.

Depending on the attack pattern, malware includes a broad range of threats or attacks such as:

- Polymorphic and metamorphic attack
- Ransomware attack
- Fileless malware attack
- Advanced Persistence Threat (APT)
- Zero-day attack and much more.

### *4.1 Polymorphic and metamorphic attack*

A polymorphic attack is a stealth strategy used by malware to create an unlimited number of new, distinct types of malwares such as trojans, viruses, worms, bots, or keyloggers which modifies itself with new variations for each attack [47]. This makes it harder for anti-malware software to detect and stop the attacks. Operating systems, server applications, and network services are the main targeted exposure of this attack.

Attack execution process - Polymorphic attacks can be implemented in a variety of diverse ways, such as Exploit mutation and shellcode polymorphism. In general, a polymorphic attack has three main elements [48] which are outlined below:

*Attack vector*- An attack vector is used to exploit the vulnerability of the target's host to get malware onto the computer or to build a mutated gene. The mutation or polymorphism attempts to conceal its true purpose, and in some cases, to make it more harmful.

*Attack body*- After the flaw has been exploited, the malicious code performs the targeted damage using the shellcode polymorphism technique. Shellcode is an element of the payload used to exploit a software flaw, and it typically contains assembly instructions that enable the attacker's remote connection. The most frequent directive in shellcode is to run a shell, which allows the code to execute commands on the computer. Other common functions include creating a privileged user profile, initiating a reversing connection to the attacker's computer, and performing various destructive activities, such as memory shifting, instruction swapping, command reordering, garbage insertions, and finally encryption.

*Polymorphic Decryptor*- This section contains the program that decodes the shellcode. It decrypts the encrypted assault body and takes control of it. The Decryptor is polymorphic and can be obfuscated in a variety of ways.

Metamorphic attack- Metamorphic malware is more complex than polymorphic malware because it employs dynamic code concealment instead of the encryption key, where the code alters with each repetition of the malicious process [49]. This continual modification makes it extremely difficult for anti-virus programs to recognize, isolate, and remove this malware[50]. There are several types of polymorphic and metamorphic families such as VirLock, Locky, Cerber, Crysis, Kelihos Botnet, Beebone, etc.

### *4.2 Ransomware attack*

Ransomware is a type of virus that typically targets consumers, the public, and corporate organizations to demand ransom money from the victims. The primary purpose of ransomware is to encrypt all data on a device, making it impossible for the victim to access the data. Cybercriminals use a variety of tactics to gain access to consumers' or institutions' documents and assets to extort a ransom. These ransom demands can come in the form of payments in real-time, or the data will be permanently destroyed if the victim does not pay up [51]. The attacker accepts payments from clients in virtual currencies such as Bitcoin, making it harder to track their names and location. Cybercriminals often use email phishing and brute force attacks to get an early footing on sensitive information and then use ransomware to take control of systems [52].

Attack Execution process- Figure 4 demonstrates the five stages of the ransomware attack chain. The following are the major stages of a ransomware attack chain:

*Infection*- The first phase is when the attacker uses various attack methods to try and get malware onto a target device. This could be done through ads, hacked websites, drive-by downloads, etc.[53], [23]. Infection can occur physically or virtually to back up data and computer memories.

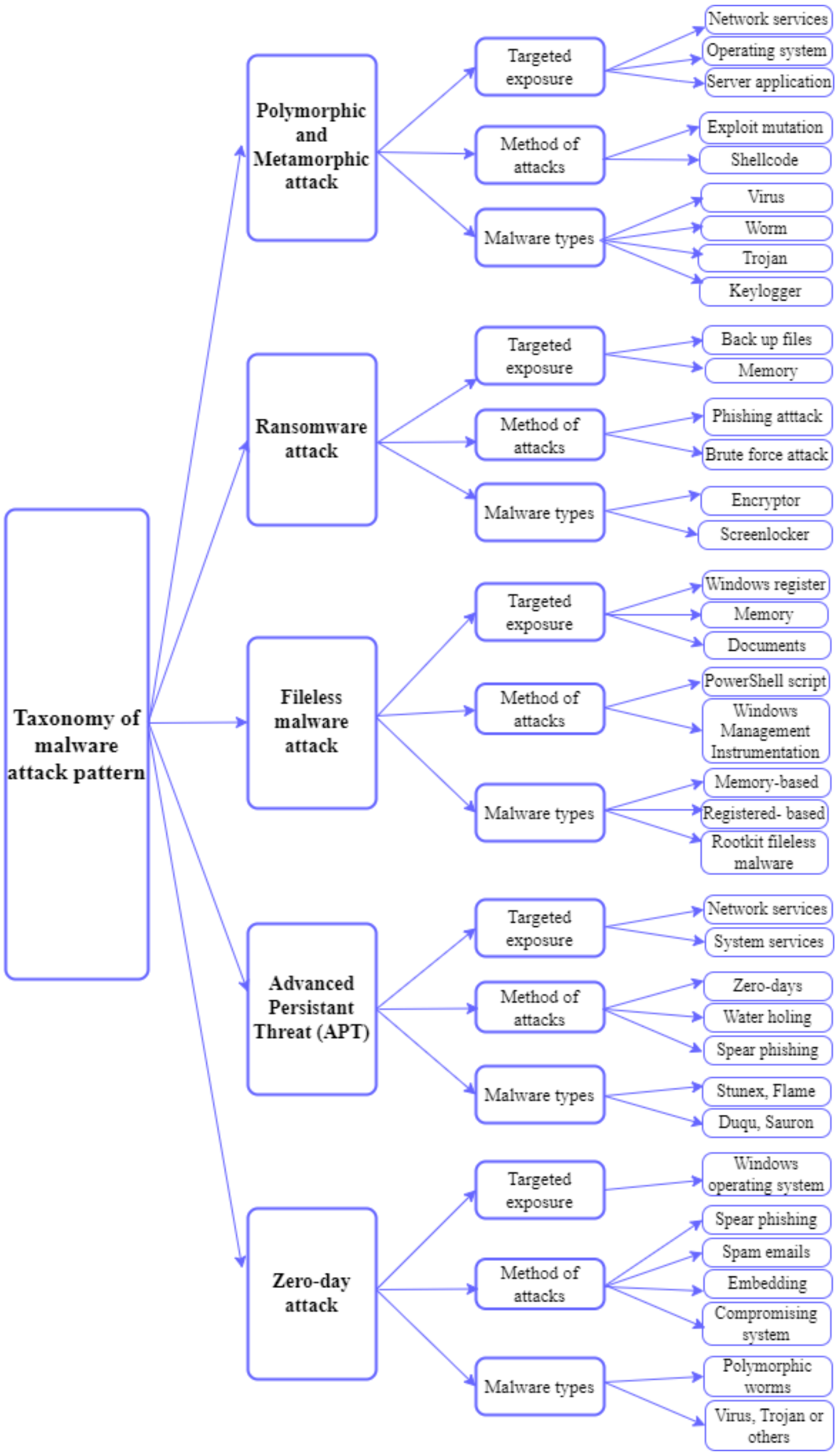

Figure 3: Taxonomy of malware attack pattern

Attack Execution process- Figure 4 demonstrates the five stages of the ransomware attack chain. The following are the major stages of a ransomware attack chain:

*Infection*- The first phase is when the attacker uses various attack methods to try and get malware onto a target device. This could be done through ads, hacked websites, drive-by downloads, etc.[53], [23]. Infection can occur physically or virtually to back up data and computer memories.

*Key generation*- Once infected, the ransomware communicates with a remote command and control server to receive instructions from the attackers on how to carry out its malicious activities. This includes retrieving a secret key and other information about the victim machine.

*Scanning*- In this phase, the malware looks for files to encrypt on the local machine and networked devices.

*Encryption*- Ransomware now conducts attacks, which include encrypting data or blocking computers to restrict the users from using their contents or computer.
*Extortion*- Finally, the ransomware shows a ransom letter to notify the victim of the attack. The ransom note reveals the facts of the attack and payment instructions. Developing picture, textual, or web pages is a typical way to keep track of notes [51],[23].

Reveton, 2012; CryptoLocker, 2013; Petya, 2016; WannaCry, 2017, PureLocker, 2018; LockerGoga, 2019; Corona ransomware, 2020; RaaS Ransomware, 2021 are some examples of ransomware families that occurred in the corresponding years.

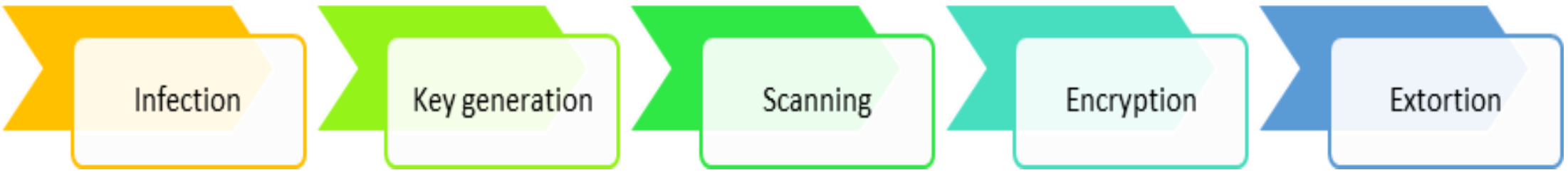

Figure 4: Ransomware attack chain

### *4.3 Fileless malware attack*

Fileless malware does not require the traditional form of malicious executables to be placed on the victim's system. This payload is delivered through alternative means such as command scripts (e.g., JavaScript, PowerShell, batch commands) or Remote Desktop Protocol (RDP) connections [54]. Fileless malware is so named because it does not produce extra files like standard malware that uses files to infect a host, instead, it inserts malicious software into the main memory. Hence, it is also called memory-based malware. As a result, fileless malware attacks are extremely effective and successful [55]. An updated report disclosed that fileless malware infections increased by 888% in 2020 [56]. The memory, documents, and Windows registry are all potential targets for fileless malware. The attacker aims to gain access to sensitive data by exploiting vulnerabilities in the unencrypted version of the victim organization's software. They use two widely used Windows tools or programs - scripts and installation processes - to inject malicious code into memory without being detected [57] namely: PowerShell scripts and Windows Management Instrumentation (WMI).

PowerShell is a program that can be used to translate simple text files into commands for Windows. Since it has access to all the files on a computer, malicious PowerShell operations are difficult to detect. The Windows Management Instrumentation (WMI) is another native tool that can be applied to launch fileless assaults. WMI is used to pass instructions to PowerShell [58]. SQL Slammer, Kovter, Phase Bot, Poweliks, Lurk Trojan, etc., are some examples of fileless malware families.

Attack execution process-Figure 5 shows the three stages attack chain of fileless malware which are as follows:
*The entry phase*-To begins, the attacker uses a variety of attack vectors, such as an infected system, a suspicious URL, a vulnerable webpage, or a malignant attachment in a malicious or spoofing email, in which the hacker tells their victims to click on a link or an attachment to obtain access to their system.

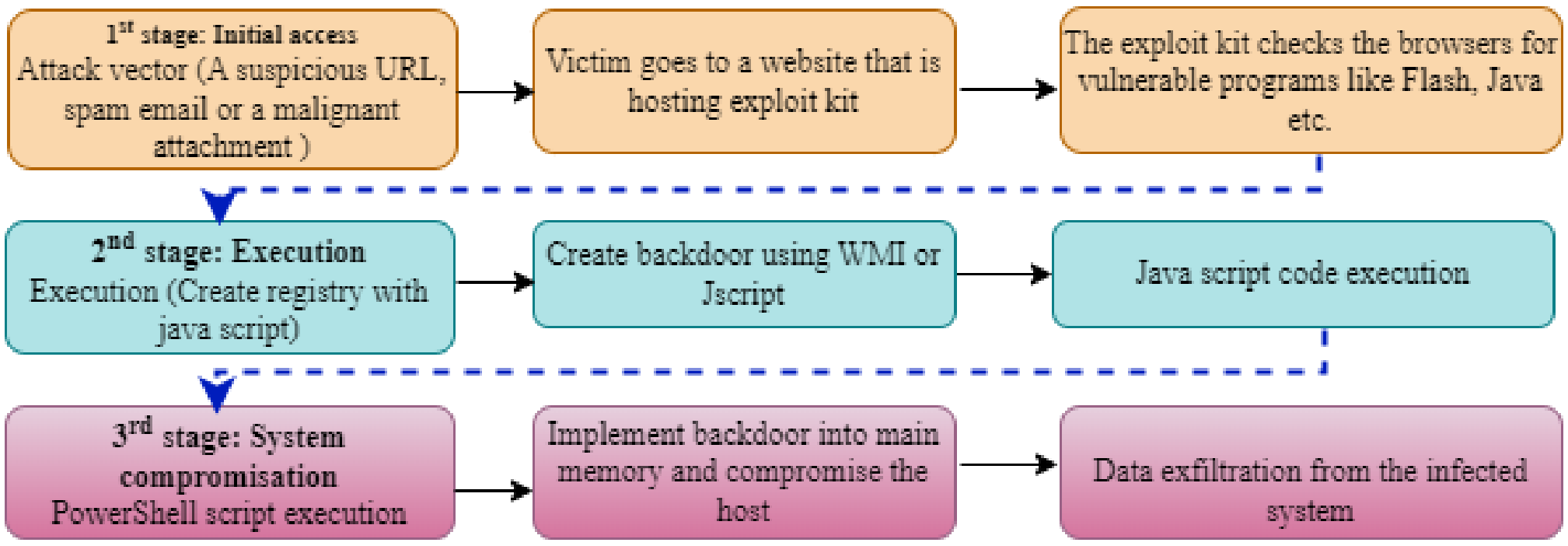

Figure 5: Attack chain stages of fileless malware

*Execution*- Second, malicious code can try to stay alive by creating plugins or using WMI and Jscript to create a backdoor. Alternatively, the code could install malicious scripts directly into memory to stay active.
*Compromised system to exfiltrate data*-Third, PowerShell can be used to install and implement malware or backdoors into memory without leaving any traces on the computer, which can be used to hijack data from targets [28].

## *4.4 Advanced Persistent Threat (APT)*

An advanced persistent threat (APT) is an attack that uses persistent malware (e.g., Stuxnet, Flame, Duqu, and Project Sauron) to gain access to a system or network over an extended period [26]. APT malware is usually built to last a long time, thus the label "persistent" [59]. It mainly focuses on stealing data from an organization, rather than damaging the system or network [60]. The majority APTs are built for specific hacking attacks and leverage sophisticated attacking methodologies such as zero-days and social engineering like Water holing and Spear phishing, spam email, etc.
*Attack execution process*- Figure 6 represents the five stages of Advance Persistence Threat (APT) to perform the data-stealing process [54],[61] which are as follows:
*Infiltration*- In the first phase, cybercriminals most commonly access their victim's computer system using social engineering, spear phishing, or zero-day malware.
*Installation*- The attacker implant malware and certain other remote management tools on the victim's computer, which allows them to control the computer remotely.
*Expansion*- At this stage, cybercriminals have gained direct control over other workspaces, server software, and network components, which gives them access to login details, such as account usernames and passwords, to gain access to crucial business information.
*Encryption*- In this phase, the attackers steal resources and information from the victim's computer, encrypting and compressing it for future exfiltration.
*Exfiltration*- Finally, the cyber attacker exfiltrates the stolen data from the victim's network. They'll then try to remove any forensic proof of the data transmission.

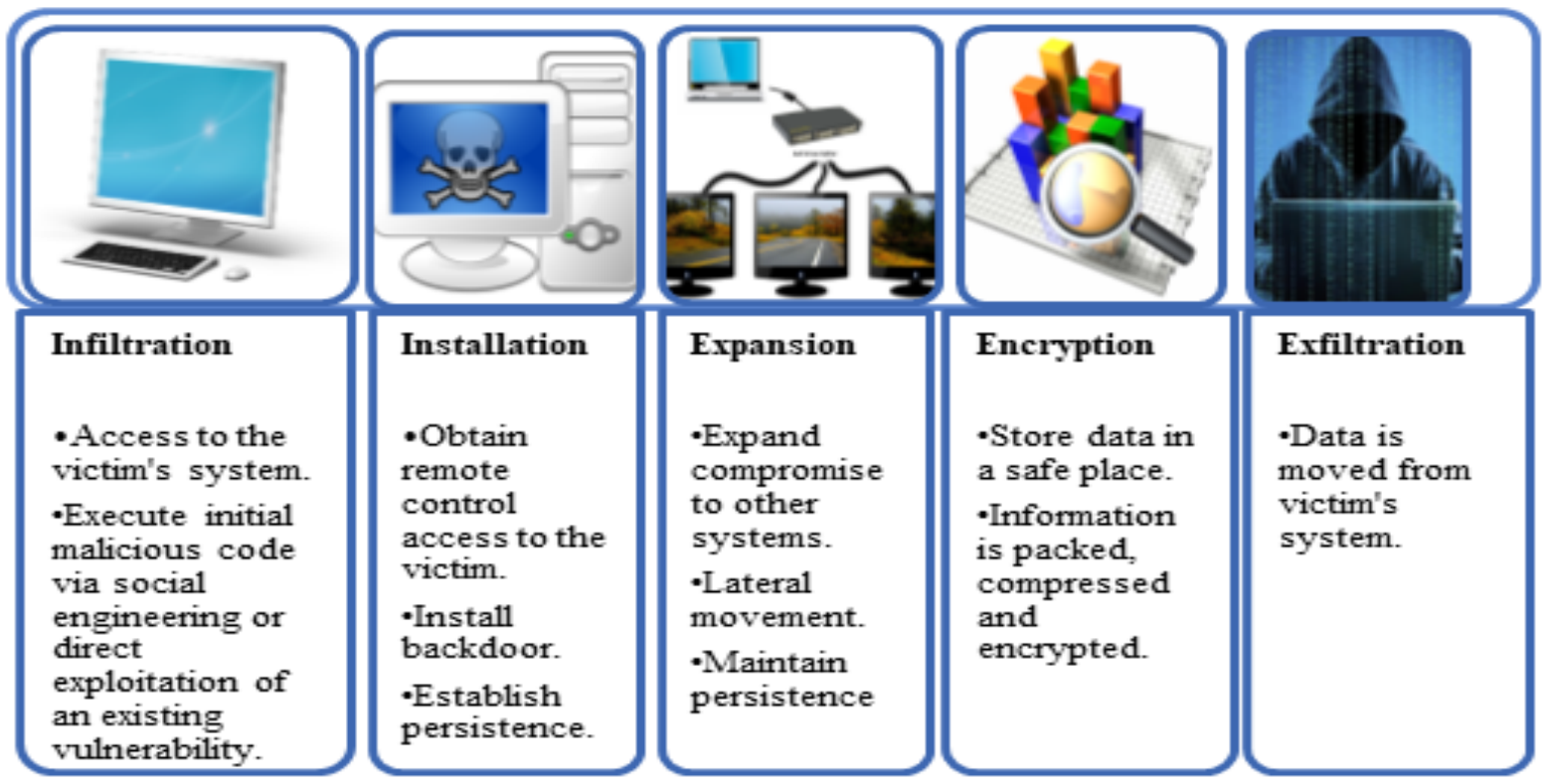

Figure 6: Stages of an Advanced Persistence Threat (APT)

## *4.5 Zero-day attack*

A zero-day attack is a type of cyberattack that uses vulnerabilities in operating systems and presents a severe hazard to internet security. Polymorphic worms, viruses, Trojans, and other malware can be used in zero-day attacks [62]. Using this method, attackers can take useful info, such as legal papers and company information. Before a weakness in hardware and software is patched, zero-day attacks are carried out. No fingerprints are left behind by zero-day assaults, making them impossible to identify [63]. Cyber attackers apply various methods for launching and executing zero-day attacks [64] including:

- Spear phishing with social engineering

- Spam emails and phishing
- Implanting exploit tools in advertisements and malicious sites
- Infecting a computer, networks, or servers.

Attack execution process- Poor computer or security settings, anti-virus software faults, and programming mistakes by professionals are the targets for zero-day attacks. Figure 7 demonstrates a zero-day attack that performs its execution in four phases [65], which are discussed below:

*Searching vulnerabilities*- Criminals look for vulnerabilities in software to exploit them, and hackers look for ways to attack essential systems or users before the developers can fix the problem.

*Exploiting*- One way to exploit a web browser vulnerability is to send emails to people, trying to trick them into visiting websites that are infected with viruses or other malware.

*Launching attack*- The zero-day exploit is a vulnerability that a hostile party has discovered, which can be used to launch attacks. The malware used in these attacks is usually complicated to detect.

*Execution and exfiltration*-The zero-day assault is executed once the zero-day exploit has been installed onto devices. In this phase, malware can harvest sensitive information like user credentials and passwords, destroy data, and ultimately take control of the computer.

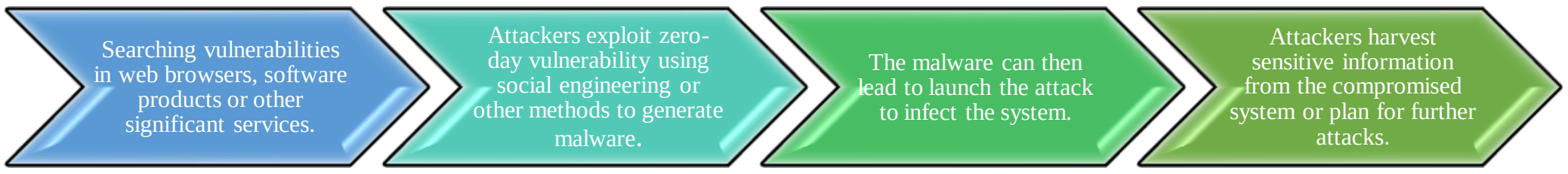

Figure 7: Zero-day attack chain

## 5. Malware detection

Identifying threats or malware by scanning computers and other files is called malware detection. Malware detection consists of several phases to identify and categorize malware. First, malware analysis is performed statically or dynamically to check whether the suspected file is malicious. Then, features are gathered. Following that, feature selection and representation are completed. Lastly, the malware classifiers are trained using classification methods. A schematic representation of the malware detection process employing ML methods is shown in Figure 8.

### *5.1 Malware analysis*

The key objective of malware analysis is to check whether a specified file or system is malevolent [66]. Three significant approaches perform malware analysis namely- static analysis, dynamic analysis, and hybrid analysis.

#### *5.1.1 Static analysis*

Static malware analysis doesn't involve running the code. It uses signatures (series of bytes) to identify malware [67]. A static analysis examines the static characteristics of a suspected file. This includes characters, passwords, signatures, and information. Signature-based detection approaches are widely preferred in cyberspace due to their simplicity, user-friendliness, low false-positive rates, and minimal processing complexity [12]. However, they require more human interaction and can detect only known malware. Different disassemblers are used to convert the malicious or binary files into assembly code such WinDbg, IDA Pro, capstone, and Ollydbg [68]. Table 4 gives a quick summary of static analysis tools.

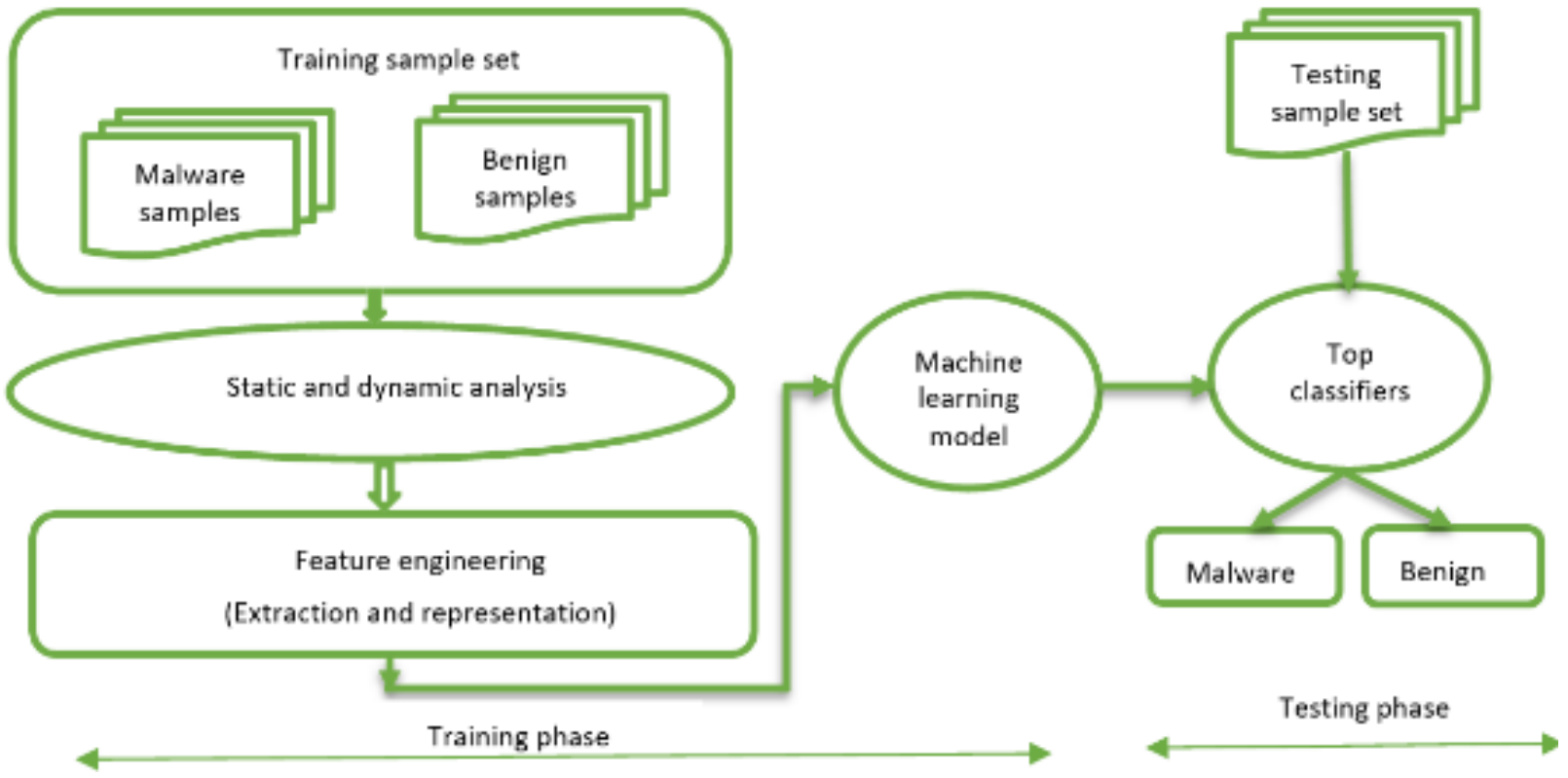

Figure 8: Schematic representation of the malware detection process employing ML methods.

Table 4: Static analysis tools.

| Name of the tool | Explanation |
|---|---|
| IDA Pro [69] | IDA Pro displays information about malicious software, which can help identify hackers. |
| PeView [70] | This tool provides information about the operating system files, including the headers that identify different types of software. The PE header information is then used in malware analysis to differentiate between harmful and malignant programs. |
| Yara [71] | The Yara tool can be used to identify strings in executable files that may be indicative of malicious behavior. |
| PEid [72] | This tool can determine if the malware is encrypted and, if so, which packer toolkit was used to encrypt it. |
| Radare [73] | Radare is a toolkit that can be used to reverse engineer various types of software, including Linux, Android, Windows, and macOS. |
| IOC Finder [74] | Indicator of Compromise (IOC) gives details on the computers that have been hacked. It creates a log report in MS Word or Web page format with details on a particular host's network and device data. |
| OllyDump [75] | OllyDump is a tool for extracting code from the system database. This method is beneficial for analyzing packaged binaries that are hard to deconstruct. |
| CFF Explorer [76] | This program displays detailed information about the executable file. |

### 5.1.2 *Dynamic analysis*

Dynamic analysis is an effective tool for observing the behaviour of a program at run-time, identifying errors and anomalies that would be difficult to detect in static code review [77]. This method generates a behaviour report on the malware-infected file's behaviours, such as its interface with the network, registry, and file system. To avoid damage to the host Windows OS, this technique conducts the study in a distinct virtual system. After the execution, it keeps track of specific features like instructions, API calls, and system calls for individual executables. HookMe and Microsoft's Detours techniques in a sandbox are used to generate the logs of the run-time behaviour of malware. The log profiles are taken outside the sandbox to further process and calculate the frequencies and parameters of API calls [78].

Monitoring techniques for dynamic analysis- Analysis techniques are a type of investigation procedure that can be used in a specific tool [79]. The following techniques are used in the run-time behavioral study [19]:

*Function call analysis*- All processes rely on function calls to execute their functions. A basic instruction that uses a function by calling its name is known as a function call. The hooking mechanism can be used to grab function calls.

*Execution control*- After the malware is run, it should check for updates to see if the malware has changed or if the operating system has changed. If there is a problem with the malware, it can be stopped before it does any damage using various techniques including debugging, binary instrumentation, stealth breakpoints, and so on [19].

*Information flow tracking*- Behavioral analysis tools use a technique called information flow tracking to record the flow of data inside malware during operation. This is also known as taint analysis because it uses tainted data [80].

*Capturing the traced report*- Tracing is a way to analyze the behavior of malware after it's been executed, and it can provide lots of useful information.

Monitoring tools for dynamic analysis- Various monitoring tools and different control environments are used to perform the techniques mentioned above and better understand the trace file. Table 5 provides a short explanation of dynamic malware investigation tools.

Table 5: Tools for dynamic analysis

| **Tools** | **Description** |
|---|---|
| Process Explorer [14] | Process Explorer is a useful application for monitoring and managing processes on a Windows system, providing precise information on the system's running activities. |
| Capture [14] | Capture is an analysis tool that uses three monitors to provide supplementary data about system activity including the file system, the registry, and the processes monitor respectively. It is a system memory or kernel-mode program. |
| VAMPiRE [81] | VAMPiRE is a tool used to halt malware execution occasionally and observe the behavior of malware and the operating system. |
| Wireshark, Tshark [82] | Wireshark and Tshark are used to examine network traffic. They can capture all incoming and outgoing packets between the computer and other devices on the network. |
| Vis [83] | Vis is a tool to trace the report left by the malicious process using the volatile memory acquisition technique that changes the OS slightly to dodge detection by malware. |
| Regshot [84] | The Regshot utility is employed to log the registry modifications produced by the executing sample. |
| TQana [85] | TQana is a framework capable of detecting a malicious Internet Explorer browser that helps to study the dynamic behavior of spyware. |
| Memoryze [86] | Memoryze is a computer forensics tool that runs on commands. It can handle the entire memory dump. |
| TCPview [87] | A type of Windows networking device that displays information about all a computer's UDP and TCP connections. |
| ApateDNS [88] | This device allows the analyst to collect DNS queries performed by malware without being instructed. It spoofs DNS replies for a certain IP address. |
| Sandboxes [89] | A sandbox is a software tool that can be used to analyze malware. It employs a variety of static and dynamic analysis approaches to produce a complete report. Different sandboxes include Cuckoo, CwSandbox, Anubis, GFI Sandbox, Parsa, etc. Additionally, sandboxes are often combined with other tools to retrieve various attributes from malware such as tcp dump for network activity and volatile for memory dumping. |

### *5.1.3 Hybrid analysis*

The hybrid approach combines dynamic and static investigation elements to produce a more detailed understanding of malicious files. To provide full disassembled explanations and supplemental strings/API call sequences, it combines real-time information with in-depth static analysis of code dumping.

## *5.2 Feature extraction*

The first step in malware detection is to obtain malicious software files. This is done through static and dynamic analysis of executable files [90]. Feature extraction is used to transform large, ambiguous data into a feature set that impacts the system's productivity, resilience, and precision. Figure 9 displays the taxonomy of feature extraction techniques. The feature extraction process is classified into two categories:

- Static approach and

- Dynamic approach

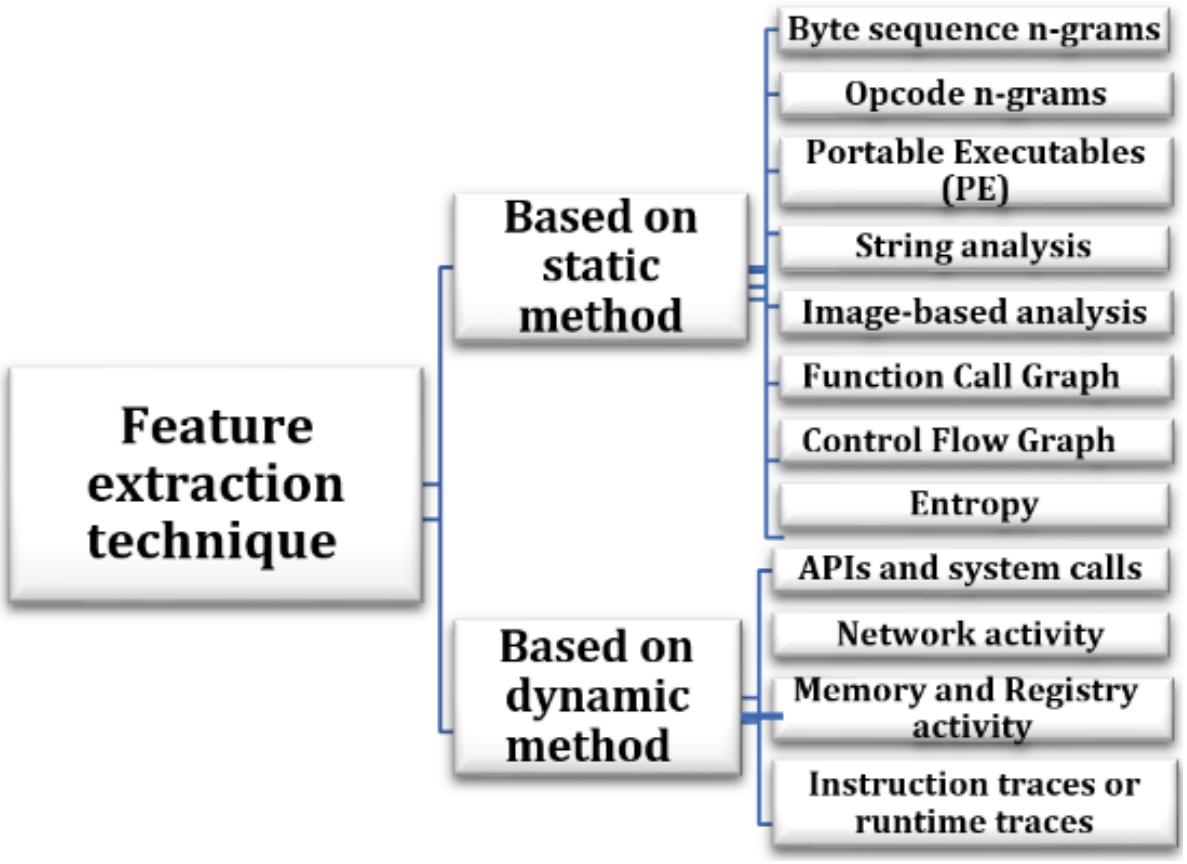

Figure 9: Taxonomy of feature extraction methods

Static feature extraction method- Static analysis is used to extract static features from binary files. These features can be used to identify malicious behavior. Two data sets are used to gather this information: the logic configuration of the application program or the machine language file recovered by changing and deconstructing the application code. There are several approaches to extracting features from evaluation articles to aid in malware detection as follows:

*Byte sequence n-grams model*- N-grams are a common technique for creating sequences of bytes from binary files. They can also be defined as byte codes and can be included in an executable file's characteristics, code, or information. Many researchers have used this approach to improve malware detection and categorization accuracy. For example, Saxe and Berlin's (2015) [91] study found that their algorithm using byte sequences was able to accurately identify malware with a 95% accuracy rate, and 0.1% FPR. Nataraj et al. (2011) [92] found that malware from the same family often shares similar visual characteristics, which can be detected using their technique. Yajamanam et al. (2018) [93] added an investigation into this, and the obtained accuracy was 92%. The same technique was applied by Bhodia et al. (2019) [94][85], where the binary is transformed into images, and found that it was very accurate. In several aspects, this study broadens and improves the strategy implemented by Yajamanam et al. For instance, the performance is compared between image-based and non-image-based analysis. Lin et al. (2015) [95] suggested a genetic approach that used about 790,000 n-grams to classify malware and obtained 90% efficiency. Furthermore, many other works rely on n-gram features for classifying malware [96], [97], [98], [99], [100],[101]. However, in most circumstances, byte sequences are unreliable.

*Opcode n-gram features*- Opcodes are short pieces of code that tell a computer what to do. They are like machine code, but they are easier to understand and can be preprocessed to provide extra information about a program (such as its name). Malware scripts are often locked up so that it's hard to figure out the sequence of bytes that make up the code, but opcodes make this easier [13]. An opcode performs numerical, analytical, and manipulation of data and can be used to determine the difference between malicious and genuine software. For example, Shabtai et al. (2012) [102] suggested an opcode n-gram feature-based malware detection framework, with n extending from 1 to 6. Anderson et al. (2012) [97] also use the transition matrix from one opcode to the other as a feature. Moreover, Santos et al. (2013) [103] used normalized opcode n-gram frequency bands to characterize executables. According to Yuxin et al. (2019) [104], the malware was identified using a deep learning algorithm that relied on static patterns and bytecode. Also, malware can be classified using static analyses by studying the opcode in the articles [101], [100], [105], [106], [107].

*Portable Executables (PE)*- The features of the PE method can be collected from the metadata saved in PE file types on a Windows system, and static analysis of PE structural information can help determine whether a file has been tampered with or corrupted to carry out malicious actions. To tackle encrypted malware, Wang and Wu (2011) [72] proposed a general Packing Detection Framework (PDF) that can be used to analyze Portable Executable (PE) files to determine if they are compressed or packed and was successful in 94.5% of packing detection cases based on analyzing 3784 non-packed executables and 1056 packed executables. The study by Kim et al. (2016) [108] found that using PE headers as features improved classification performance. A PE's static analysis can offer much useful information, including sectors, importers, keywords, and compilers [109], [110], [91], [111].

*String analysis*- String features are retrieved from executables using clear text from program files such as window frames, message boxes, get versions, libraries, etc. These strings are readable characters encrypted in PE and non-PE executable code. Static analysis of a PE can be used to check for strings, including code fragments, creator signatures, data types, and system-relevant data [100], [91]. Printable strings are binary features that indicate whether a string is present in an executable. Dahl et al. (2013) [112] and Huang and Stokes (2016) [113] extracted uninitialized objects spilled from pictures of a folder in storage as usable characters. Islam et al. (2013) [114] used the string program in IDA Pro5 to get understandable words or strings from the entire file. However, most malicious programs do not depend on printed strings to perform tasks.

*Image-based analysis*- Nataraj et al. (2011) [92] pioneered a technique for visualizing malware binaries, by converting every value into a digital image with counts 0-255. This digital image is then used to characterize a malware image. The technique is approximately 40 times faster than traditional methods, and its accuracy is 98 percent. Furthermore, Le et al. (2018) [115] found a malicious program that can transform an entire executable into a sequence of images. The method was tested on a large sample of binaries and yielded an accuracy of 98.8%. A similar technique was used by Bhodia et al. (2019) [94] and looked at ways to improve the strategy used by Yajamanam et al. (2018) by comparing its performance between image-based and non-image-based analyses.

*Function Call Graph (FCG)*- A graph of function calls within a software program is created by static analysis. IDA Pro or Radare2 can later retrieve this information [33]. Some articles use the function call feature to detect and classify malware. For example, Kinable et al. (2011) [116] proposed a way to identify malicious code based on the structural similarities between its function call graphs. The method was used to compare the graphs of 1050 different malware samples. Islam et al. (2013) [114] found that the length of an executable's code can be used to identify different malware variants. They counted the sum of code bytes in executables to get this information. Furthermore, Hassen and Chan (2017) [117] suggested a rapid and accurate malware classification approach based on feature vector extraction from the Function Call Graph model achieves an overall accuracy of 0.979 on a smaller dataset.

*Control Flow Graph*- A graph of the program's flow is composed of nodes, which represent system calls and API calls. The control flow graph is used to represent the program's behavior and to identify the relationships between the different parts of the program [66]. Control flow graphs are used in various articles to identify malicious software. For example, Eskandari and Hashemi (2011) [118] described a technique for detecting metamorphic malware using control flow graphs and achieved 97% accuracy in identifying these types of malware. Later, Faruki et al. (2012) [119] looked at PE files to see if there were any abnormal patterns in the API calls, they contained They then used a CFG to reconstruct API calls and used n-gram processing to transform these API calls into input vectors ranging from 1 to 4.

*Entropy*-The entropy of a byte's series measures its numerical variation in terms of information theory concepts. Zero entropy indicates that the same characters have been reused throughout the code block, while a byte with high entropy contains many different values. To spot malware, Sorokin and Jun (2011) [120] investigated how entropy varies among folders by comparing them to a training set. Baysa et al. (2013) [121] improved on earlier research by using wavelet methods to detect places where the entropy levels changed significantly. In addition, the paper by Wojnowicz et al. (2016) [122] calculated the entropy of a document's wavelet-based energy distribution, and then used a variety of logistic regression models to see how much of a change in entropy would make the document seem malicious.04

Dynamic (run-time) feature extraction method- Dynamically analyzed function calls or behavior data types are stated as dynamic features like APIs and system calls, function parameter analysis, Control Flow graphs (CFG), network activity, registry activity, and file system which are as follows:

*APIs and system calls*- APIs and system calls represent malware behavior, and most run-time behavioral analyses trust the usage of API calls as the key feature in identifying malicious process activities [123], [100], [85]. In addition, using an emulator, Bai et al. (2014) [110] and Santos et al. (2013) [103] extracted the API calls method dynamically. Similarly, by performing an executable in a virtual machine, Islam et al. (2013) [114] and Dahl et al. (2013) [112], Uppal et al. (2014) [124] captured API function calls and associated variables for malware classification. In addition, a behavior-based model was projected by Galal et al. (2015) [125], Ki et al. (2015) [126], Liang et al. (2016) [127], and Xiaofeng et al. (2019) [128] to identify the malware's run-time activities. However, the combined model outperformed the separate models by 96.7 percent. Few other authors employed system calls as features to investigate the malware samples. For example, the study by Kolosnjaji et al. (2016) [129] looked at how to find out which file systems or operating systems were initiated by an application program. Moreover, Anderson et al. (2012) [97] and Huang and Stokes (2016) [113] classify operating systems into large groups, each representing a functionally related set of system operations, including display paintings or file writing. The same features are used to classify the malware in other papers, e.g., Elhadi et al. (2013) [130] and Mao et al. (2015) [131].

*Network activity*- Monitoring the PE's communication with the networks can provide useful information, such as how to communicate with a central server. Information on used networks, TCP/UDP channels, database queries, and DNS-level interactions can all be useful tools. Many reviewed articles collected this type of information using dynamically extracted network activities as a feature set [132], [133], [127], [134]. In addition, Bekerman et al. (2015) [135] developed a framework that analyzes traffic on the network to identify malicious programs. Arivudainambi et al. (2019) [136] focused on building a model that was used to detect malware samples from network artifacts and achieved 99% accuracy. In addition, a Probabilistic Neural Network (PNN) framework was proposed by Rabbani et al. (2020) [137] for identifying malicious activity in network attacks. which resulted in a 96.5% detection accuracy.

*Memory and Registry activity*- At runtime, the contents of the computer's main memory can be used to infer the behavior of a computer program. In addition, the data saved in various registers during execution can provide valuable information about the context of a program, and it is one of the most important ways for a program to communicate with the Windows OS. Ghiasi et al. (2015) [138] proposed a method for detecting behavior similarity between two sets of data by analyzing the memory and registering data. Yucel et al. (2020) [139] also developed strategies for capturing malicious activities based on executable file memory pictures. Liu (2020) [140] proposed a way to mitigate the effects of Adversarial Examples (AE) using adversarial training and data visualization. Vasan (2020) [141] also proposed an ensemble of CNN models to classify malware based on images. Furthermore, the research of Singh and Singh (2020) [32] and Escudero García and DeCastro-García, (2021) [142] focused on behavior-based malware detection methods using optimum feature sets. The researchers detected malware with 99.54 % and 98% accuracy.

*Instruction traces or run-time traces*-A dynamic run-time trace is a sequence of CPU commands that are executed while the code is running. This differs from static instruction traces, which are arranged as they exist in the binary format. Anderson et al. (2011) [97] proposed an infection recognition system relying on examining maps built from the code tracing acquired during the targeted executable's run-time. Again, Storlie et al. (2014) [143] demonstrated malicious file findings depending on automatically generated instruction traces investigation. Carlin et al. (2017a) [144] also described a method for extracting program run-time traces both from legitimate and malignant executable files using dynamic response on virtual computers. Moreover, Ali et al. (2017) [89] developed a machine-learning algorithm to identify malicious files, and their results showed that it was 99% accurate. Alaeiyan et al. (2019) [145] proposed a method for detecting anomalies that relied on run-time variables.

### *5.3 ML-based malware classification*

There has been increased interest in using machine learning techniques to predict and categorize malware over the last decade. A workflow in an ML model is an ongoing procedure that includes obtaining available data, purifying, and formatting it, constructing models, evaluating them, and putting them into operations. Figure 10 shows the workflow of machine learning that illustrates how the ML model works for automatic detection and classification.

Numerous machine learning classifiers were used to train the system in the literature. A classification algorithm and the training data or selected features develop a machine learning model. A quick comparison of the different ML classifiers is presented in Table 6.

Table 6: Different machine learning algorithms, their strengths, and weaknesses.

| ML algorithms | Key idea | Strength | Weakness |
|---|---|---|---|
| SVM (Support Vector Machine) [146] | An SVM is a popular machine learning algorithm used to classify data such as "malignant" and "benign." This approach relies on finding the best set of points that divides the data into these two groups. | Provides better classification accuracy and can handle large datasets and nonlinear patterns. | The training time in SVM becomes very long. |
| Decision Tree (DT) [147] | A DT algorithm is a tool that can categorize data by grouping it into tree-like structures based on the information gained from each feature. The feature with the most information gain becomes the root of the tree, and other features become leaves. | DT classifiers can handle high-dimensional datasets and inconsistent data.<br><br>DT classifiers are effective at quickly classifying data. | The model may not be reliable if the data is not accurately represented. |
| Naive Bayes (NB) [148] | The probability of a particular classification is determined by using the NB technique for both single and multi-classifications. | The NB sorting technique is easy to follow and understand. | It performs poorly when the training datasets are related. |
| Artificial Neural Network (ANN) [146] | A neural network is a technology that tries to imitate how humans learn. It has nodes that are linked together, and three different layers are used to represent these nodes. | The non-linear dataset can be modeled using ANN, which can be used to solve practically any problem. | Overfitting may be an issue for ANN and functionally limited. |
| K-Nearest Neighbor (K- NN) [149] | An unsupervised learning algorithm is used to distinguish between two data instances. | It is quicker to train and more space-resistant than other classifiers. | Its computation time is an overhead. |
| Random Forest (RF) [150] | Random Forest is an ensemble learning algorithm that generates a series of decision trees from a subset of the training dataset. | Less risk of overfitting, hence better results even if the dataset changes. | Slow training speed and high costs. |
| Logistic Regression (LR) [133] | LR is a binary classification approach that is parameterized by the qualitative responses of the data sets. This determines the variables' experimental values. | Less complicated, simpler to analyze, and can deal with non-linear effects. | The average prediction accuracy is poor. |
| Boosting and Bagging Algorithms [89] | Boosting algorithms can be used to train multiple weak classifiers together, and a classification model can be built by combining the results. Alternatively, a different form of ensemble learning known as bagging can be used. | Every sort of data is well served by boosting methods.<br><br>Applied to train the weak classifiers. | Boosting techniques consume much time and thus are computationally costly. |

This section provides a survey of the various machine-learning algorithms used by analyzed articles to identify and categorize malware based on its various attributes.

Nataraj et al. (2011) [92], Le et al. (2018) [115], and Bhodia et al. (2019) [94] all offered methods for malware families classification utilizing neural networks and grayscale images as static features, but somewhat distinct classifiers were used for malware classification. In this regard, Nataraj et al. proposed a K-NN classifier to identify malware images and detected 98% of images from 25 families. The Convolutional Neural Network (CNN) algorithm was applied by Le et al. to predict malicious programs using 10568 binary data and was found to be 98.8% accurate. Bhodia et al. compared the outcomes of natural image-dependent deep learning (DL) to a basic K-NN technique and found that a basic K-NN technique performed better than deep learning when it came to classifying malware.

Euh (2020) [151] proposed an approach to identify malicious programs using a tree-based ensemble model depending on n-grams of bytes and opcodes, APIs, and a WEM (Window Entropy Map). The same feature extraction techniques as n-gram of byte and opcode sequence were used in other studies Lin et al. (2015) [95], Shabtai et al. (2012) [102], Santos et al. 2013 [103], Raff et al. (2018) [152], and Yuxin et al. (2019) [104] but these are distinct from each other in terms of using different classifiers and the achieved accuracy levels. For example, Santos et al. (2013) developed an n-grams-based file signature using an SVM classifier. The study by Shabtai et al. (2012) looked at eight different classifiers and found that the best performance was achieved with a 96% accuracy rate and a 0.1% false-positive rate. Yuxin et al., (2019), used DT, SVM, and the K-NN algorithms as classifiers, and found that the DT algorithm was the most effective, with an overall 95.8% accuracy.

Kim et al. (2016) [108] described an approach that uses static features like PE headers and ML classifiers like SVM and Stochastic Gradient Descent (SGD) to detect malware. It was 99 percent accurate and had a 0.2 percent false-positive rate. Nagano and Uda (2017) [153] also suggested a mechanism to spot malware with the help of K-NN and SVM classifiers which yielded an efficiency of 99%. Besides these, different classifiers were used in different papers using static features. For example, Raff et al. (2017) [99] used Logistic Regression, Wang and Wu (2011) [72] used SVM, Faruki et al. (2012) [119] used RT, DT, NB, and Elhadi et al. (2013) [130] employed Graph Matching Algorithm to notice harmful threats.

Galal et al. (2015) [125] and Ki et al. (2015) [126] presented a behavior-based strategy to identify run-time malware by analyzing API call sequences. Several classification techniques, for example, SVM, RF, and DT algorithms were used in the study by Galal et al., and the Multiple Sequence Alignment algorithm (MSA), and Longest Common Subsequence (LCS) techniques were used by Ki et al. to detect malware. Their achieved accuracy level was 97.19% and 99.8% respectively. In addition, Xiaofeng et al. (2019) [128] used the API call technique to extract the features and RF algorithm to train and classify the malicious program. Mohaisen et al. (2015) [133], Pektas Acarman (2017) [154], Singh and Singh (2020) [32], and Escudero García and DeCastro-García (2021) [142] presented different approaches to classify malware families that relied on different attributes like network, registry, and file system activities applying different ML algorithms. In this case, Singh and Singh (2020) used ensemble machine learning methods and attained the greatest level of accuracy of 99.54%. Pektaş and Acarman (2017) developed online machine-learning algorithms (CW, ARROW, PA-I & II, NHERD) to classify malware. Hyperparameter optimization algorithms (Bayesian and Random type) were used by Escudero García and DeCastro-García (2021) and the accuracy was greater than 98%. Arivudainambi et al. (2019) [136] focused on building a model for malware detection utilizing Neural Network (NN), Principal Component Analysis (PCA), and Convolutional neural network (CNN with an accuracy of 99%. In addition, a Probabilistic Neural Network (PNN) framework was suggested by Rabbani et al. (2020) [137] for identifying malicious activity in network attacks which resulted in a 96.5% performance measure.

Namavar et al. (2020) [155] applied the ensemble learning approach based on behavioral features and have a 99.65% accuracy rate. In addition, Vasan, (2020) [141] and Damaševičius (2021) [156] used ensemble learning and achieved an accuracy of 99% and 99.99 % respectively. But Amer and Zelinka (2020) [157] used Markov chain representation and K-means. Their detection performance was 99.9 % and an FP rate of 0.010.

A malware classification system is also based on a hybrid framework in the articles by Islam et al. (2013) [114], Shijo et al. (2015) [158], Mangialardo et al. (2015) [159], Kolosnjaji et al. (2016) [129], Ali et al. (2017) [89], Huda et al. (2018) [78], Kumar et al. (2019) [160], and Gupta & Rani (2020) [161], Damodaran et al. (2017) [162], Han et al. (2019b) [163] that relied on dissimilar ML algorithms like SVM, RF, DT, IB1, CNN, XGBoost, K-NN, etc. The maximum precision obtained from most of the experimental findings was more than 99%. For example, the study of Huda et al. (2018) used multiple classifiers such as MR+SVM, Fisher+SVM, and MRED+SVM and demonstrated an accuracy of 99.49%. Kumar et al. (2019) obtained 99.74 % accuracy for the proposed hybrid strategy, and Gupta and Rani (2020) achieved a 99.5% accuracy rate using a variety of classifiers, including Neural Networks, Random Forests, Decision Trees, and XGBoost.

## 6. Comparative analysis and findings

This section provides a comparative analysis of different methods for detecting malware, with a focus on developing an effective and novel machine-learning model with the lowest false-positive rate. To select such a model, we have done extensive literature reviews to find the best methods and summarized what is known about them. Table 7 shows which malware detection algorithms are most used, as well as the variety of malware features that are used.

In addition, Table 8 presents the summary of the presented papers based on some important factors such as malware analysis methods, feature extraction strategies, and ML classifiers. Figure 14 shows the proportion of data analysis techniques that have been used in studies, with dynamic analysis being the most popular (46%). The hybrid analysis came in second (29%), followed by static analysis (25%). Figure 15 depicts how well different malware detection methods work. The SVM (23%) is the most successful, followed by RF (18%), DT (16%), KNN (14%), Boost (9%), NB (5%), and NN (4%). The SMV (static-based) strategy is the most accurate.

Table 7: A side-by-side comparison of the most recent reviewed papers on malware detection based on feature types, number of samples, classification models, and accuracy [K-NN = k-nearest neighbor, Classification and Regression Tree = CART, Stochastic Gradient Descent = SGD, Random Forest = RF, logical regression = LRA, Principal Component Analysis = PCA, Support Vector Machine = SVM, Decision Tree = DT].

| Authors | Feature types | No. of samples (M = Malware, B = Benign) | Classification algorithms | Accuracy |
|---|---|---|---|---|
| The static-based malware detection approach | | | | |
| Le et al. (2018) [115] | Grayscale image | M = 10568 | CNN | 98.8% |
| Bhodia et al. (2019) [94] | Grayscale image | – | K-NN, DT | 99.60% |
| Lin et al. (2015) [95] | N-gram features | M = 4288 | SVM | 95% |
| Yuxin et al. (2019) [104] | N-grams opcode | M= 400 | DBN, SVM, and k−means | 95.8% |
| Euh (2020) [151] | N-gram of bytes and opcodes, APIs, and WEM | M = 122,963 | XGBoost, random forest, AdaBoost, extra trees | 98.5% |
| Kim et al. (2016) [108] | PE headers | M = 27,000 and B = 11,000 | SVM, CART, and SGD | 99% |
| Nagano and Uda (2017) [153] | DLL import, assembly code, and hex dump. | M = 3600 | SVM and K-NN | 99% |
| Eskandari and Hashemi (2011) [119] | Control Flow Graphs | M = 2140, B = 2305 | RF | 97 %. |
| Ahmadi et al. (2016) [100] | Strings, opcode, API function calls, frequency of keywords | M = 21741 | XGBoost, Gradient Boosting | 99.76 % |
| Hassen and Chan (2017) [117] | Function Call Graph model | Small dataset | Minha's signature | 97.9% |
| The dynamic-based malware detection approach | | | | |
| Galal et al. (2015) [125] | API call sequences | M = 2000, B = 2000 | DT, RF, SVM | 97.19% |
| Ki et al. (2015) [126] | API call sequences | M = 23080 | MSA and LCS | 99.8 % |
| Xiaofeng et al. (2019) [128] | API call sequences | M = 1430, B = 1352 | RF and RNN | 96.7% |
| Mohaisen et al. (2015) [136] | API calls, network, registry, memory, and system files. | M = (400−115,000) | SVM, logical regression, and hierarchical clustering | 98% |
| Singh and Singh (2020) [32] | API calls, PSI, file operations, registry, and network activities. | M=16489, B= 8422 | Ensemble machine learning methods | 99.54 % |
| Escudero García and DeCastro-García, (2021) [142] | API calls, network traffic, file system, and registry. | M = 9999, B = 9995 | Bayesian and Random type optimization algorithm. | 98% |
| Ali et al. (2017) [89] | Run-time features | M = 150000, B = 87000 | SVM, DT, and Boosted DT | 99% |
| Namavar et al. (2020) [155] | Behavioral features | M = 18831 | Ensemble learning | 99.65% |
| Arivudainambi et al. (2019) [136] | Network artifacts | M= 1000 | PCA, NN, and CNN | 99%. |
| Amer and Zelinka (2020) [157] | API calls and System calls | M = 30658, B = 21422 | Markov chain representation, K−means | 99.9%, |
| Vasan (2020) [141] | Natural image | M = 9339 | Ensemble CNN | 99% |
| Damaševičius (2021) [156] | DOS image header, File header features, packer, and DLL. | M = 2722, B = 2488 | Ensemble CNN, SVM, RF, DT, KNN, AdaBoost, etc. | 99.99% |
| Hybrid−based malware detection approaches | | | | |
| Islam et al. (2013) [114] | Integrated feature set (FLF, PSI, API calls) | M= 2939, B=541 | SVM, DT, RF, IB1 | 95% |
| Shijo et al. file (2015) [158] | Printable string information (PSI)APIs | M= 997, B=490 | SVM, RF | 98.7% |
| Damodaran et al. (2017) [162] | PE Header Information, API calls | M=745, B= 40 | Hidden Markov Model (HMM) | 98% |
| Huda et al. (2018) [78] | Run time features | M = 2000, B = 1500 | MR+SVM, Fisher+SVM, MRED+SVM | 99.49% |
| Han et al. (2019b) [163] | PE files, API calls, DNS, and DLL information. | M= 4250 | K-NN, RF, DT, and Extreme Gradient Boosting | 97.21% |
| Kumar et al. (2019) [160] | PE metadata, network data, system calls, process, and registry | 120000-item dataset | RF, DT, Neural Network, XGBoost, and K-NN | 99.74% |
| Gupta & Rani (2020) [161] | Metadata, packers, API calls, File system, Network & Registry activity | M = 100,200, B = 98,150 | Ensemble learning, RF, SVM, DT, NB, KNN | 99.5%. |

Furthermore, the different features used by different malware classification tools can impact malware detection accuracy. Figure 16 demonstrates the feature selection process in three types of malware detection techniques. Some researchers prefer to use specific or single features to develop a malware detection technique, while others use multiple features. The study found that the author Ki et al. used single API calls to achieve the highest accuracy of 99.8%. Ahmadi et al. achieved the highest accuracy of 99.76 % using multiple static features, while Damaševičius achieved the highest accuracy using dynamic features.

This study looked at the accuracy of malware detection methods and found that most of the recent studies have used SVM with other classifiers and ensemble learning models to achieve great accuracy levels as shown in Figures 17 and 18 respectively. It was also found that a hybrid malware detection method with ML methods can be more effective than using just one type of detection method. For this technique, Kumar et al. achieved a maximum accuracy of 99.74 % as shown in Figure 19.

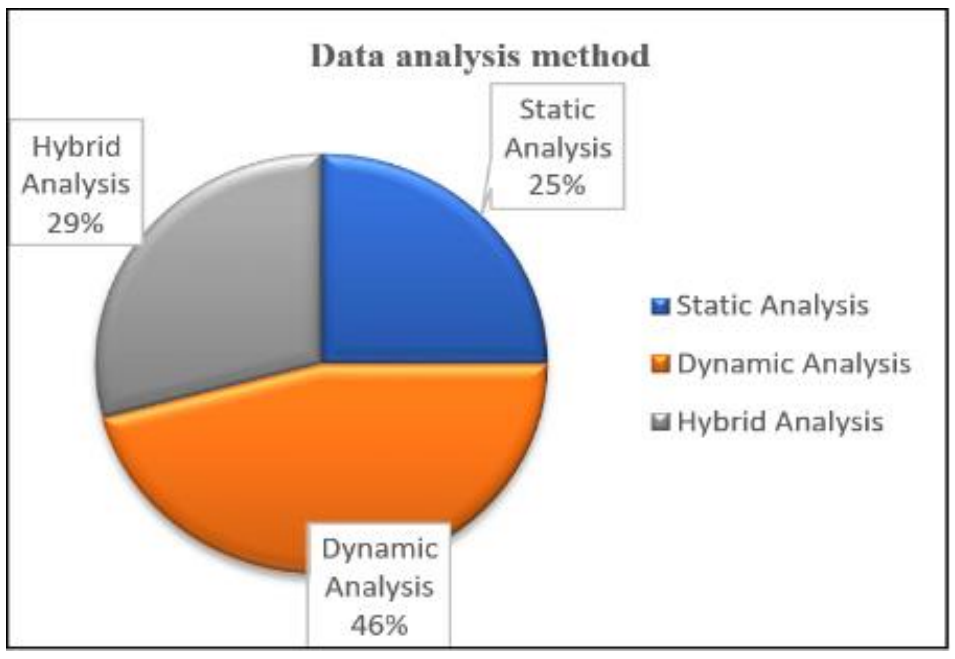

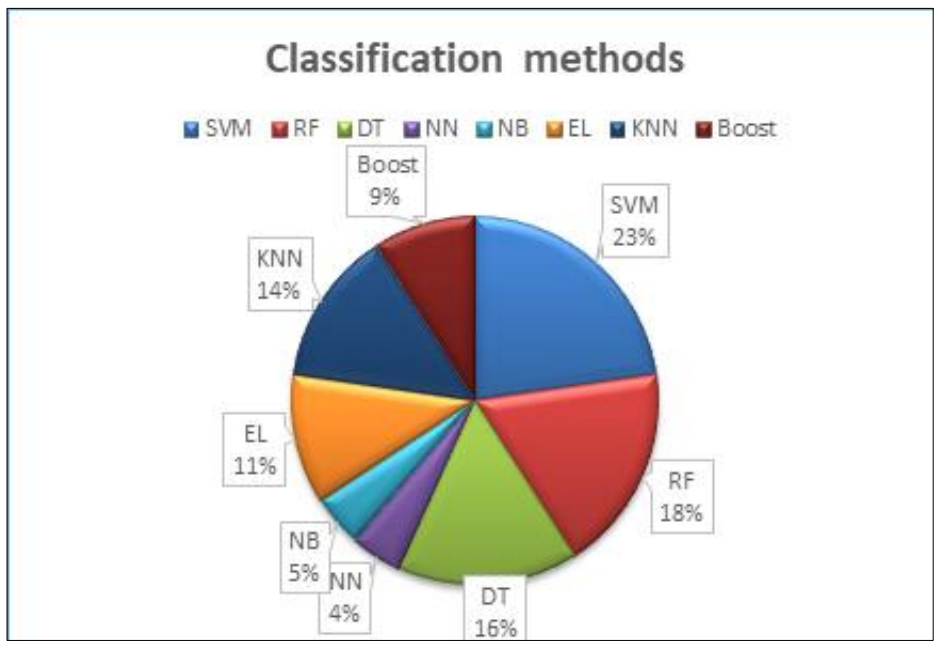

Figure 14: Comparison of analysis techniques Figure 15: Comparison of ML algorithms

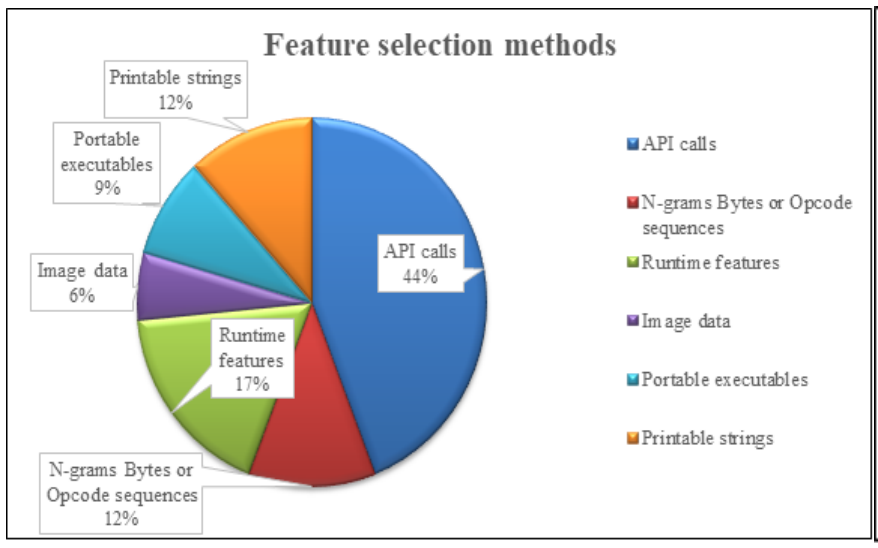

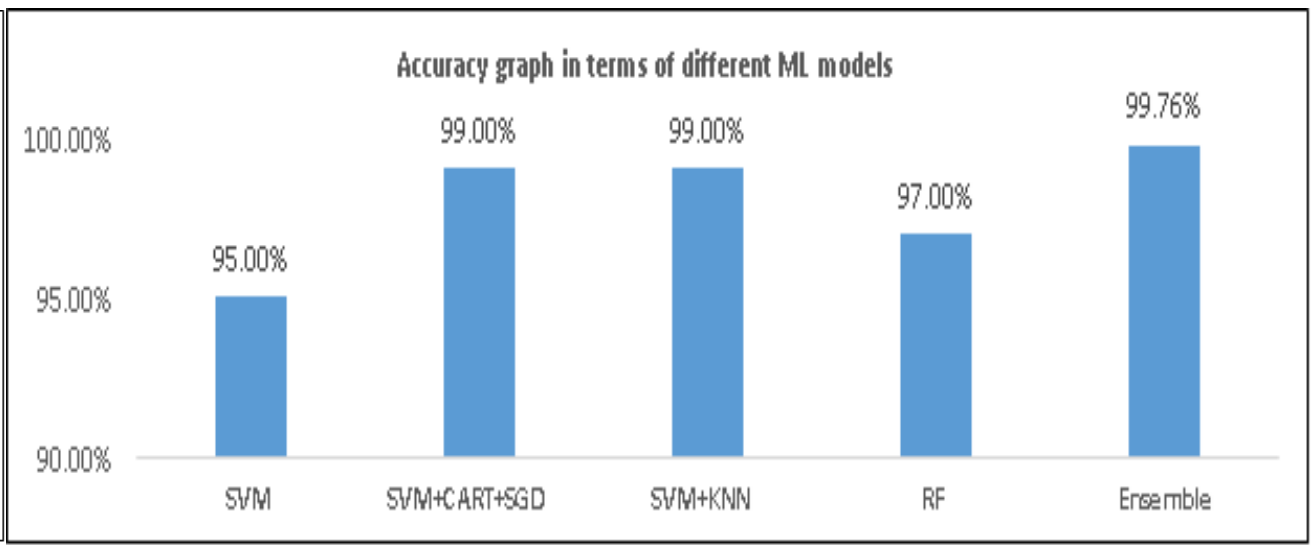

Figure 16: Comparison of feature selection methods Figure 17: Accuracy comparing chart in static methods

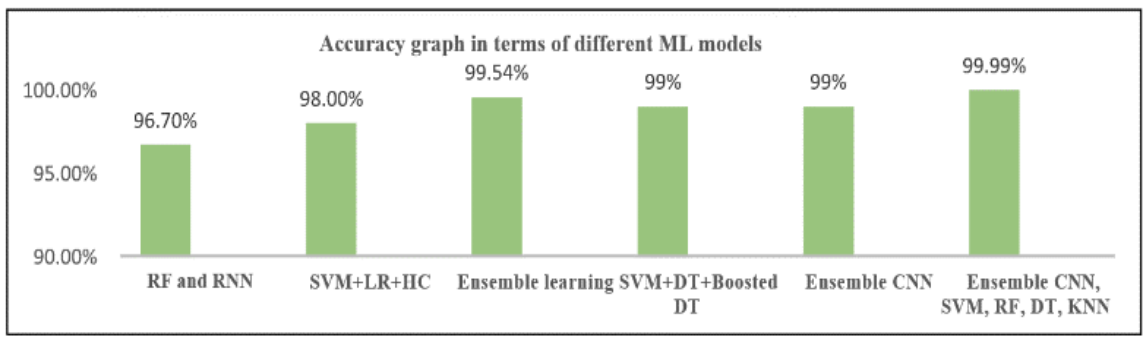

Figure 18: Accuracy comparing chart in dynamic malware detection methods

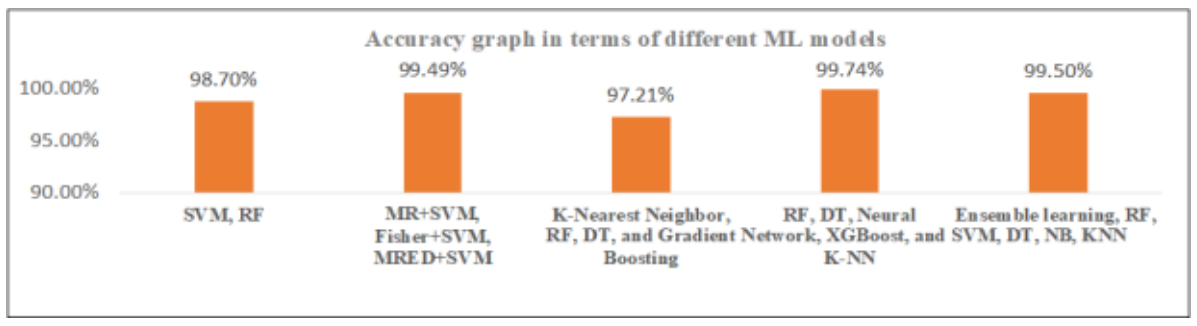

Figure 19: Accuracy comparing chart in hybrid malware detection techniques

Table 8: Summary of the most recent reviewed papers on malware detection based on three types of analysis, machine learning algorithms, and malware features [SA = Static Analysis, DA = Dynamic Analysis, SVM = Support Vector Machine, RF = Random Forest, DT = Decision Trees, NN = Neural Network, CNN =Convolutional Neural Network, LR = Logistic Regression, NB = Naïve Bayes, K-NN = k-Nearest Neighbour, NBO = N-grams Bytes or Opcode sequences, CFG = (Control flow Graph, Function call Graph), ID = Image Data, PEH = Portable Executable Header, PSI = Printable String Information, RTF = Runtime Features (File, Network, Resistivity), ASI = (API calls, System calls, Imports, DLL Imports]

| Authors | Analysis types | | | ML algorithms | | | | | | | | | | Feature types | | | | | | |
|---|---|---|---|---|---|---|---|---|---|---|---|---|---|---|---|---|---|---|---|---|
| | SA | DA | HA | SVM | RF | DT | NN | LR | NB | RNN | EL | KNN | BOOST | NBO | CFG | ID | PEH | ASI | PSI | RTF |
| Kim et al. (2016) [108] | • | | | • | | | | | | | | | | | | | • | | | |
| Yuxin et al. (2019) [104] | • | | | • | | | | | | | | | | • | | | | • | | |
| Xiaofeng et al. (2019) [128] | | • | | | • | | | | | • | | | | | | | | • | | |
| Bhodia et al. (2019) [94] | • | | | | | | | | | | | • | • | • | | | | • | | |
| Ali et al. (2017) [89] | | • | | • | | • | | | | | | | | | | | | | | • |
| Ahmadi et al. (2016) [100] | • | | | | | | | | | | | | • | • | | | | • | • | |
| Escudero García and DeCastro- García (2021) [142] | | • | | | • | | | | • | | | | | | | | | • | | • |
| Namavar et al. (2020) [155] | | • | | | | | | | | | • | | | | | | | | | • |
| Arivudainambi et al. (2019) [136] | | • | | | | | • | | | | | | | | | | | | | • |
| Nagano and Uda (2017) [153] | • | | | • | | | | | | | | • | | | | | | • | • | |
| Singh and Singh (2020) [34] | | • | | | | | | | | | • | | | | | | | • | | • |
| Amer and Zelinka, (2020) [157] | | • | | | | | | | | | | • | | | | | | • | | |
| Vasan, (2020) [141] | | • | | | | | | | | | • | | | | | • | | | | |
| Damaševičius, (2021) [156] | | • | | • | • | • | | | | | • | • | | | | • | • | • | | |
| Islam et al. (2013) [114] | | | • | • | • | • | | | | | | | | | | | | • | • | • |
| Kim et al. (2016) [108] | | • | | • | | • | | | | | | | | | | | | • | | |
| Mohaisen et al. (2015) [136] | | • | | • | | | | • | | | | | | | | | | | | |
| Han et al. (2019b) [163] | | | • | | • | • | | | | | | | | | | | • | • | | |
| Kumar et al. (2019) [160] | | | • | | • | • | • | | | | | • | • | | | | | • | | • |
| Gupta and Rani (2020) [161] | | | • | • | • | • | | | • | | • | • | | | | | | • | | • |
| Huda et al. (2018) [78] | | | • | • | | | | | | | | | | | | | | • | | |
| Euh (2020) [151] | • | | | | • | | | | | | | | • | | | | | | • | |

Finally, it has been found that SVM and API calls are the most used classifiers and feature types for static, dynamic, and hybrid malware detection methods. Furthermore, a single classification technique is ineffective for constructing a model with multiple malware features. Nevertheless, SVM outperformed the others in static analysis. Ensemble methods also performed exceptionally well during dynamic and hybrid analysis. Hence, the ensemble of algorithms would be most appropriate for finding and classifying malware in terms of accuracy and economic analysis. This method also can attain more effective feature selection to overcome encryption issues.

## 7. Research issues and challenges

Malware detection is a never-ending procedure. It's getting more difficult day by day. As the number of computer users grows, criminals are building sophisticated malicious activities that are hard to notice. Along with the sophistication of malware, the rate at which it is generated is a huge concern in preventing malware attacks. As a result, the conflict between malicious actors and experts becomes more intense as technology advances. However, after analyzing various malware detection approaches, some major research concerns or limitations have been discovered. Some specific challenges in malware detection that remain unsolved include:

- Real-time monitoring is a continuing contest. Several recent studies have used data for detecting anomalous files that are not ideal for monitoring.
- The parameterization of used algorithms is another important component in the effectiveness of malware classifiers. It is a significant contributor to the malware classifier's accuracy. This issue is not treated in-depth in the proposed method that has been reviewed.
- Malware attackers create adversarial strategies to compel the classification model to mislead the training data (e.g., by feeding it incorrect data). The model should understand adversarial tactics, resulting in a more efficient and reliable detection scheme.
- Most malware detection methods are vulnerable to False-Positive Rates (FPs) and False Negative Rates (FNs). Some features and signatures in malicious and benign files can be similar, raising FPs and FNs.
- ML algorithms are sensitive to overfitting and bias in practice. This results in lower Detection Rates (DRs) and higher FPs

## 8. Conclusion and future work

When sensitive data is shared through hyper-connected networks, a run-time malware attack may compromise data privacy. To protect data from malware threats, this paper presents an exhaustive literature review of attack pattern taxonomy and machine learning-based malware detection and classification methods. These different types of attacks pattern are presented broadly in section 4. This provides information on how devices can be compromised, and how to protect them. In addition, the malware detection and classification techniques have been discussed in section 5. A summary of the most recent reviewed papers on malware detection based on some important factors is also presented in Table 7 and Table 8. According to the findings, SVM and API calls are extensively used in classification models and feature types respectively for static, dynamic, and hybrid malware attack detection. We found that the SVM method has the highest accuracy among machine learning-based static malware detection approaches. However, when building a model with multiple malware features, a single classification technique is unproductive. During dynamic and hybrid analysis, ensemble methods consistently performed well. Moreover, Table 1 demonstrates the summary and comparison of our works with others. Finally, this paper explores the most pressing research dilemmas that researchers face.

However, focusing future research on the parameterization of utilized algorithms could be beneficial in malware classification effectiveness. This issue is not explored in full in the proposed method. In addition, special emphasis on real-time malware identification and attack protection on a vast dataset to avoid scalability may develop future research on accurate malware detection because existing solutions only focus on a limited amount of data.

## Acknowledgments

This research study was supported by the CSCRC (Cyber Security Cooperative Research Centre Limited), which is funded partially by the Australian Government's Cooperative Research Centre's Program.

## References

[1] J. Jang-Jaccard and S. Nepal, "A survey of emerging threats in cybersecurity," *J. Comput. Syst. Sci.*, vol. 80, no. 5, pp. 973–993, Aug. 2014, doi: 10.1016/j.jcss.2014.02.005.

[2] FORTINET, "Cybersecurity Statistics," 2022, [Online]. Available: https://www.fortinet.com/resources/cyberglossary/cybersecurity-statistics.

[3] Devon, "15 Alarming Cyber Security Facts and Stats," 2020. https://www.cybintsolutions.com/cyber-security-facts-stats/ (accessed Dec. 23, 2022).

[4] M. Souppaya and K. Scarfone, "NIST Special Publication 800-83 Revision 1 - Guide to Malware Incident Prevention and Handling for Desktops and Laptops," *NIST Spec. Publ.*, vol. 800, p. 83, 2013, doi: 10.6028/NIST.SP.800-83r1.

[5] O. Aslan and R. Samet, "A Comprehensive Review on Malware Detection Approaches," *IEEE Access*, vol. 8, pp. 6249–6271, 2020, doi: 10.1109/ACCESS.2019.2963724.

[6] L. Caviglione *et al.*, "Tight Arms Race: Overview of Current Malware Threats and Trends in Their Detection," *IEEE Access*, vol. 9, pp. 5371–5396, 2021, doi: 10.1109/ACCESS.2020.3048319.

[7] A. Souri and R. Hosseini, "A state-of-the-art survey of malware detection approaches using data mining techniques," *Human-centric Comput. Inf. Sci.*, vol. 8, no. 1, 2018, doi: 10.1186/s13673-018-0125-x.

[8] Z. Bazrafshan, H. Hashemi, S. M. H. Fard, and A. Hamzeh, "A survey on heuristic malware detection techniques," *IKT 2013 - 2013 5th Conf. Inf. Knowl. Technol.*, pp. 113–120, 2013, doi: 10.1109/IKT.2013.6620049.

[9] I. Basu, N. Sinha, D. Bhagat, and S. Goswami, "Malware Detection Based on Source Data using Data Mining : A Survey," *Am. J. Adv. Comput.*, vol. III, no. 1, pp. 18–37, 2016.

[10] J. Singh and J. Singh, "A survey on machine learning-based malware detection in executable files," *J. Syst. Archit.*, vol. 112, p. 101861, Jan. 2021, doi: 10.1016/J.SYSARC.2020.101861.

[11] R. Sihwail, K. Omar, and K. A. Z. Ariffin, "A survey on malware analysis techniques: Static, dynamic, hybrid and memory analysis," *Int. J. Adv. Sci. Eng. Inf. Technol.*, vol. 8, no. 4–2, pp. 1662–1671, 2018, doi: 10.18517/ijaseit.8.4-2.6827.

[12] S. Abijah Roseline and S. Geetha, "A comprehensive survey of tools and techniques mitigating computer and mobile malware attacks," *Comput. Electr. Eng.*, vol. 92, p. 107143, Jun. 2021, doi: 10.1016/j.compeleceng.2021.107143.

[13] A. Abusitta, M. Q. Li, and B. C. M. Fung, "Malware classification and composition analysis: A survey of recent developments," *J. Inf. Secur. Appl.*, vol. 59, p. 102828, Jun. 2021, doi: 10.1016/j.jisa.2021.102828.

[14] E. Gandotra, D. Bansal, S. Sofat, E. Gandotra, D. Bansal, and S. Sofat, "Malware Analysis and Classification: A Survey," *J. Inf. Secur.*, vol. 5, no. 2, pp. 56–64, Feb. 2014, doi: 10.4236/JIS.2014.52006.

[15] S. Singla, E. Gandotra, D. Bansal, and S. Sofat, "Detecting and Classifying Morphed Malwares: A Survey," *Int. J. Comput. Appl.*, vol. 122, no. 10, pp. 28–33, 2015, doi: 10.5120/21738-4937.

[16] B. Yu, Y. Fang, Q. Yang, Y. Tang, and L. Liu, "A survey of malware behavior description and analysis," *Front. Inf. Technol. Electron. Eng. 2018 195*, vol. 19, no. 5, pp. 583–603, Jul. 2018, doi: 10.1631/FITEE.1601745.

[17] S. Choudhary and A. Sharma, "Malware Detection Classification using Machine Learning," *Proc. - 2020 Int. Conf. Emerg. Trends Commun. Control Comput. ICONC3 2020*, Feb. 2020, doi: 10.1109/ICONC345789.2020.9117547.

[18] M. Egele, T. Scholte, E. Kirda, and C. Kruegel, "A survey on automated dynamic malware-analysis techniques and tools," *ACM Comput. Surv.*, vol. 44, no. 2, 2012, doi: 10.1145/2089125.2089126.

[19] O. Or-Meir, N. Nissim, Y. Elovici, and L. Rokach, "Dynamic malware analysis in the modern era—A state of the art survey," *ACM Comput. Surv.*, vol. 52, no. 5, 2019, doi: 10.1145/3329786.

[20] S. Talukder and Z. Talukder, "A Survey on Malware Detection and Analysis Tools," *Int. J. Netw. Secur. Its Appl.*, vol. 12, no. 2, pp. 37–57, 2020, doi: 10.5121/ijnsa.2020.12203.

[21] S. Abijah Roseline and S. Geetha, "A comprehensive survey of tools and techniques mitigating computer and mobile malware attacks," *Comput. Electr. Eng.*, vol. 92, p. 107143, Jun. 2021, doi: 10.1016/J.COMPELECENG.2021.107143.

[22] E. Berrueta, D. Morato, E. Magana, and M. Izal, "A Survey on Detection Techniques for Cryptographic Ransomware," *IEEE Access*, vol. 7, pp. 144925–144944, 2019, doi: 10.1109/ACCESS.2019.2945839.

[23] H. Oz, A. Aris, A. Levi, and A. S. Uluagac, "A Survey on Ransomware: Evolution, Taxonomy, and Defense Solutions," *ACM Comput. Surv.*, Feb. 2022, doi: 10.1145/3514229.

[24] R. Moussaileb, N. Cuppens, J. L. Lanet, and H. Le Bouder, "A Survey on Windows-based Ransomware Taxonomy and Detection Mechanisms," *ACM Comput. Surv.*, vol. 54, no. 6, Jul. 2021, doi: 10.1145/3453153.

[25] A. Sharma and S. K. Sahay, "Evolution and Detection of Polymorphic and Metamorphic Malwares: A Survey," *Int. J. Comput. Appl.*, vol. 90, no. 2, pp. 7–11, 2014, doi: 10.5120/15544-4098.

[26] S. Sibi Chakkaravarthy, D. Sangeetha, and V. Vaidehi, "A Survey on malware analysis and mitigation techniques," *Computer Science Review*, vol. 32. pp. 1–23, 2019, doi: 10.1016/j.cosrev.2019.01.002.

[27] R. Kaur and M. Singh, "A survey on zero-day polymorphic worm detection techniques," *IEEE Commun. Surv. Tutorials*, vol. 16, no. 3, pp. 1520–1549, 2014, doi: 10.1109/SURV.2014.022714.00160.

[28] Sudhakar and S. Kumar, "An emerging threat Fileless malware: a survey and research challenges," *Cybersecurity*, vol. 3, no. 1, pp. 1–12, 2020, doi: 10.1186/s42400-019-0043-x.

[29] Mary Landesman, "A Brief History of Malware."

[30] J. R. & I. Belcic, "What Is Malware? The Ultimate Guide to Malware." https://www.avg.com/en/signal/what-is-malware.

[31] J. Singh and J. Singh, "A survey on machine learning-based malware detection in executable files," *J. Syst. Archit.*, vol. 112, no. March 2020, p. 101861, 2021, doi: 10.1016/j.sysarc.2020.101861.

[32] J. Singh and J. Singh, "Detection of malicious software by analyzing the behavioral artifacts using machine learning algorithms," *Inf. Softw. Technol.*, vol. 121, p. 106273, May 2020, doi: 10.1016/J.INFSOF.2020.106273.

[33] D. Gibert, C. Mateu, and J. Planes, "The rise of machine learning for detection and classification of malware: Research developments, trends and challenges," *J. Netw. Comput. Appl.*, vol. 153, p. 102526, Mar. 2020, doi: 10.1016/J.JNCA.2019.102526.

[34] ClickSSL, "Identify Sources Of Malware – How To Handle?," 2017. https://www.clickssl.net/blog/identify-sources-of-malware-how-to-handle (accessed May 08, 2022).

[35] M. A. Royi Ronen, Marian Radu, Corina Feuerstein, Elad Yom-Tov, "Microsoft Malware Classification Challenge," 2018, [Online]. Available: https://arxiv.org/abs/1802.10135.

[36] H. S. Anderson and P. Roth, "EMBER: An Open Dataset for Training Static PE Malware Machine Learning Models," 2018, [Online]. Available: http://arxiv.org/abs/1804.04637.

[37] "No Title." https://virusshare.com/.

[38] "MalShare," [Online]. Available: https://www.malshare.com/about.php.

[39] "SOREL-20M," [Online]. Available: https://github.com/sophos-ai/SOREL-20M.

[40] "Windows Malware Dataset with PE API Calls," [Online]. Available: https://www.kaggle.com/datasets/focatak/malapi2019.

[41] "MalwareBazaar Database," [Online]. Available: https://bazaar.abuse.ch/browse/.

[42] "InQuest / malware-samples," [Online]. Available: https://github.com/InQuest/malware-samples.

[43] Mila, "Contagio Malware Dump," 2020, [Online]. Available: http://contagiodump.blogspot.com/.

[44] "theZoo aka Malware DB," [Online]. Available: https://thezoo.morirt.com/.

[45] "VirusSign," [Online]. Available: https://www.virussign.com/index.html.

[46] Amit Sethi Sean Barnum, "Attack Pattern Generation," 2013. https://www.cisa.gov/uscert/bsi/articles/knowledge/attack-patterns/attack-pattern-generation (accessed May 14, 2022).

[47] "WHAT ARE POLYMORPHIC ATTACKS AND HOW CAN YOU DEFEND AGAINST THEM?," 2019. https://level4it.com/2019/07/24/what-are-polymorphic-attacks-and-how-can-you-defend-against-them/ (accessed Jul. 24, 2022).

[48] P. F. M. S. R. P. O. K. W. Lee, "Polymorphic Blending Attacks," 2006.

https://www.usenix.org/legacy/event/sec06/tech/full_papers/fogla/fogla_html/main.html.

[49] S. Alam, R. N. Horspool, I. Traore, and I. Sogukpinar, "A framework for metamorphic malware analysis and real-time detection," *Comput. Secur.*, vol. 48, pp. 212–233, Feb. 2015, doi: 10.1016/J.COSE.2014.10.011.

[50] J. B. Fraley and M. Figueroa, "Polymorphic malware detection using topological feature extraction with data mining," *Conf. Proc. - IEEE SOUTHEASTCON*, vol. 2016-July, 2016, doi: 10.1109/SECON.2016.7506685.

[51] T. R. Reshmi, "Information security breaches due to ransomware attacks - a systematic literature review," *Int. J. Inf. Manag. Data Insights*, vol. 1, no. 2, p. 100013, 2021, doi: 10.1016/j.jjimei.2021.100013.

[52] D. Palmer, "Ransomware: These are the two most common ways hackers get inside your network," 2021. https://www.zdnet.com/article/ransomware-these-are-the-two-most-common-ways-hackers-get-inside-your-network/ (accessed Jul. 29, 2022).

[53] T. Dargahi, A. Dehghantanha, P. N. Bahrami, M. Conti, G. Bianchi, and L. Benedetto, "A Cyber-Kill-Chain based taxonomy of crypto-ransomware features," *J. Comput. Virol. Hacking Tech.*, vol. 15, no. 4, pp. 277–305, 2019, doi: 10.1007/s11416-019-00338-7.

[54] A. Gendre, "Malware Defense: Protecting Against Metamorphic and Polymorphic Malware," *Vade Secure*, 2017. https://www.vadesecure.com/en/polymorphic-malware/ (accessed Aug. 01, 2022).

[55] "The Rise of Fileless Malware and Attack Techniques," 2021. https://ukdiss.com/examples/fileless-malware-attack-techniques.php#_Toc508123740 (accessed Dec. 23, 2022).

[56] P. Muncaster, "Fileless Malware Detections Soar 900% in 2020," 2021. https://www.infosecurity-magazine.com/news/fileless-malware-detections-soar-1/ (accessed Mar. 30, 2022).

[57] E. H. Academy, "What is a Fileless Malware Attack - Stages & How it Works," 2021. https://ethicalhackersacademy.com/blogs/ethical-hackers-academy/fileless-malware (accessed Aug. 23, 2022).

[58] S. COOPER, "Fileless malware attacks explained (with examples)," 2021. https://www.comparitech.com/blog/information-security/fileless-malware-attacks/ (accessed May 14, 2022).

[59] Wenke Lee, "Malware and Attack Technologies Knowledge Area," 2021. https://cybok.org/media/downloads/Malware_Attack_Technologies_v1.0.1.pdf.

[60] W. Tounsi and H. Rais, "A survey on technical threat intelligence in the age of sophisticated cyber attacks," *Comput. Secur.*, vol. 72, pp. 212–233, Jan. 2018, doi: 10.1016/J.COSE.2017.09.001.

[61] C. Crane, "Polymorphic Malware and Metamorphic Malware: What You Need to Know," 2021. https://www.thesslstore.com/blog/polymorphic-malware-and-metamorphic-malware-what-you-need-to-know/ (accessed Mar. 25, 2022).

[62] Illumio, "What are Zero-Day Exploits vs. Zero-Day Vulnerabilities vs. Zero-Day Attacks?," 2021.

[63] CYBERDEFENCES, "Understanding Zero-Day Attacks," 2019. https://cyberdefenses.com/understanding-zero-day-attacks/ (accessed Jan. 15, 2022).

[64] A. P. Singh, "A Study on Zero Day Malware Attack," *Ijarcce*, vol. 6, no. 1, pp. 391–392, 2017, doi: 10.17148/ijarcce.2017.6179.

[65] U. K. Singh, C. Joshi, and D. Kanellopoulos, "A framework for zero-day vulnerabilities detection and prioritization," *J. Inf. Secur. Appl.*, vol. 46, pp. 164–172, Jun. 2019, doi: 10.1016/j.jisa.2019.03.011.

[66] D. Ucci, L. Aniello, and R. Baldoni, "Survey of machine learning techniques for malware analysis," *Comput. Secur.*, vol. 81, pp. 123–147, Mar. 2019, doi: 10.1016/J.COSE.2018.11.001.

[67] N. Udayakumar, V. J. Saglani, A. V. Cupta, and T. Subbulakshmi, "Malware Classification Using Machine Learning Algorithms," *Proc. 2nd Int. Conf. Trends Electron. Informatics, ICOEI 2018*, pp. 1007–1012, Nov. 2018, doi: 10.1109/ICOEI.2018.8553780.

[68] E. Raff *et al.*, "An investigation of byte n-gram features for malware classification," *J. Comput. Virol. Hacking Tech.*, vol. 14, no. 1, pp. 1–20, Feb. 2018, doi: 10.1007/s11416-016-0283-1.

[69] C. H. Malin, E. Casey, and J. M. Aquilina, "Analysis of a Malware Specimen," *Malware Forensics F. Guid. Wind. Syst.*, pp. 363–504, Jan. 2012, doi: 10.1016/b978-1-59749-472-4.00006-8.

[70] M. Sikorski and A. Honig, "Practical Malware Analysis: The Hands-On Guide to Dissecting Malicious Software - Michael Sikorski, Andrew Honig - Google Books," 2012. https://books.google.co.uk/books?hl=en&lr=&id=DhuTduZ-pc4C&oi=fnd&pg=PR2&dq=practical+malware+analysis&ots=3fcOToP7bw&sig=hQzoEAe6V6Z_fPNMdX1MqCrURzA&redir_esc=y#v=onepage&q&f=false (accessed Feb. 23, 2022).

[71] M. Cohen, "Scanning memory with Yara," *Digit. Investig.*, vol. 20, pp. 34–43, Mar. 2017, doi: 10.1016/j.diin.2017.02.005.

[72] T. Y. Wang and C. H. Wu, "Detection of packed executables using support vector machines," *Proc. - Int. Conf. Mach. Learn. Cybern.*, vol. 2, pp. 717–722, 2011, doi: 10.1109/ICMLC.2011.6016774.

[73] S. Schrittwieser and S. Katzenbeisser, "Code obfuscation against static and dynamic reverse engineering," *Lect. Notes Comput. Sci. (including Subser. Lect. Notes Artif. Intell. Lect. Notes Bioinformatics)*, vol. 6958 LNCS, pp. 270–284, 2011, doi: 10.1007/978-3-642-

24178-9_19.

[74] X. Liao, K. Yuan, X. Wang, Z. Li, L. Xing, and R. Beyah, “Acing the IOC game: Toward automatic discovery and analysis of open-source cyber threat intelligence,” *Proc. ACM Conf. Comput. Commun. Secur.*, vol. 24-28-Octo, pp. 755–766, Oct. 2016, doi: 10.1145/2976749.2978315.

[75] T. W. York and D. MacAlister, “Healthcare Security Risks and Vulnerabilities,” *Hosp. Healthc. Secur.*, pp. 49–77, Jan. 2015, doi: 10.1016/b978-0-12-420048-7.00003-9.

[76] S. Abimannan and R. Kumaravelu, “A Mathematical Model of HMST Model on Malware Static Analysis,” *https://services.igi-global.com/resolvedoi/resolve.aspx?doi=10.4018/IJISP.2019040106*, vol. 13, no. 2, pp. 86–103, Jan. 1AD, doi: 10.4018/IJISP.2019040106.

[77] H. Bai, C. Z. Hu, X. C. Jing, N. Li, and X. Y. Wang, “Approach for malware identification using dynamic behaviour and outcome triggering,” *IET Inf. Secur.*, vol. 8, no. 2, pp. 140–151, 2014, doi: 10.1049/iet-ifs.2012.0343.

[78] S. Huda, R. Islam, J. Abawajy, J. Yearwood, M. M. Hassan, and G. Fortino, “A hybrid-multi filter-wrapper framework to identify run-time behaviour for fast malware detection,” *Futur. Gener. Comput. Syst.*, vol. 83, pp. 193–207, Jun. 2018, doi: 10.1016/j.future.2017.12.037.

[79] W. Aman, “A Framework for Analysis and Comparison of Dynamic Malware Analysis Tools,” *Int. J. Netw. Secur. Its Appl.*, vol. 6, no. 5, pp. 63–74, 2014, doi: 10.5121/ijnsa.2014.6505.

[80] N. Kaur and A. Kumar, “A Complete Dynamic Malware Analysis,” *Int. J. Comput. Appl.*, vol. 135, no. 4, pp. 20–25, 2016, doi: 10.5120/ijca2016908283.

[81] A. Vasudevan and R. Yerraballi, “Cobra: Fine-grained malware analysis using stealth localized-executions,” *Proc. - IEEE Symp. Secur. Priv.*, vol. 2006, pp. 15–29, 2006, doi: 10.1109/SP.2006.9.

[82] V. Ndatinya, Z. Xiao, V. R. Manepalli, K. Meng, and Y. Xiao, “Network forensics analysis using Wireshark,” *Int. J. Secur. Networks*, vol. 10, no. 2, pp. 91–106, 2015, doi: 10.1504/IJSN.2015.070421.

[83] M. Yu, Z. Qi, Q. Lin, X. Zhong, B. Li, and H. Guan, “Vis: Virtualization enhanced live forensics acquisition for native system,” *Digit. Investig.*, vol. 9, no. 1, pp. 22–33, 2012, doi: 10.1016/j.diin.2012.04.002.

[84] D. Gibert, C. Mateu, and J. Planes, “The rise of machine learning for detection and classification of malware: Research developments, trends and challenges,” *J. Netw. Comput. Appl.*, vol. 153, no. December 2019, p. 102526, 2020, doi: 10.1016/j.jnca.2019.102526.

[85] M. Egele, T. Scholte, E. Kirda, and C. Kruegel, “A survey on automated dynamic malware-analysis techniques and tools,” *ACM Comput. Surv.*, vol. 44, no. 2, Mar. 2008, doi: 10.1145/2089125.2089126.

[86] C. Rathnayaka and A. Jamdagni, “An efficient approach for advanced malware analysis using memory forensic technique,” *Proc. - 16th IEEE Int. Conf. Trust. Secur. Priv. Comput. Commun. 11th IEEE Int. Conf. Big Data Sci. Eng. 14th IEEE Int. Conf. Embed. Softw. Syst.*, pp. 1145–1150, 2017, doi: 10.1109/Trustcom/BigDataSE/ICESS.2017.365.

[87] Pirscoveanu RHansen SLarsen TStevanovic MPedersen JCzech A, “Analysis of Malware behavior: Type classification using machine learning,” *2015 Int. Conf. Cyber Situational Awareness, Data Anal. Assessment, CyberSA 2015*, 2015, doi: 10.1109/CYBERSA.2015.7166115.

[88] O. Aslan and R. Samet, “Investigation of possibilities to detect malware using existing tools,” *Proc. IEEE/ACS Int. Conf. Comput. Syst. Appl. AICCSA*, vol. 2017-Octob, pp. 1277–1284, Mar. 2018, doi: 10.1109/AICCSA.2017.24.

[89] Q. K. Ali Mirza, I. Awan, and M. Younas, “CloudIntell: An intelligent malware detection system,” *Futur. Gener. Comput. Syst.*, vol. 86, pp. 1042–1053, 2018, doi: 10.1016/j.future.2017.07.016.

[90] S. Euh, H. Lee, D. Kim, and D. Hwang, “Comparative analysis of low-dimensional features and tree-based ensembles for malware detection systems,” *IEEE Access*, vol. 8, pp. 76796–76808, 2020, doi: 10.1109/ACCESS.2020.2986014.

[91] J. Saxe and K. Berlin, “Deep neural network based malware detection using two dimensional binary program features,” *2015 10th Int. Conf. Malicious Unwanted Software, MALWARE 2015*, pp. 11–20, Feb. 2016, doi: 10.1109/MALWARE.2015.7413680.

[92] L. Nataraj, S. Karthikeyan, G. Jacob, and B. S. Manjunath, “Malware images: Visualization and automatic classification,” *ACM Int. Conf. Proceeding Ser.*, 2011, doi: 10.1145/2016904.2016908.

[93] S. Yajamanam, V. R. Samuel Selvin, F. Di Troia, and M. Stamp, “Deep learning versus GIST descriptors for image-based malware classification,” *ICISSP 2018 - Proc. 4th Int. Conf. Inf. Syst. Secur. Priv.*, vol. 2018-Janua, pp. 553–561, 2018, doi: 10.5220/0006685805530561.

[94] N. Bhodia, P. Prajapati, F. Di Troia, and M. Stamp, “Transfer learning for image-based malware classification,” *ICISSP 2019 - Proc. 5th Int. Conf. Inf. Syst. Secur. Priv.*, pp. 719–726, 2019, doi: 10.5220/0007701407190726.

[95] C.-T. LIN1, N.-J. WANG1, and HAN XIAO2 AND CLAUDIA ECKERT2, “Feature Selection and Extraction for Malware Classification,” *J. Inf. Sci. Eng. 3*, vol. 31, no. 3, pp. 965–992, 2015.

[96] Z. Fuyong and Z. Tiezhu, “Malware detection and classification based on n-grams attribute similarity,” *Proc. - 2017 IEEE Int. Conf.*

*Comput. Sci. Eng. IEEE/IFIP Int. Conf. Embed. Ubiquitous Comput. CSE EUC 2017*, vol. 1, pp. 793–796, Aug. 2017, doi: 10.1109/CSE-EUC.2017.157.

[97] B. Anderson, C. Storlie, and T. Lane, "Improving malware classification: Bridging the static/dynamic gap," *Proc. ACM Conf. Comput. Commun. Secur.*, pp. 3–14, 2012, doi: 10.1145/2381896.2381900.

[98] Z. Feng *et al.*, "HRS: A hybrid framework for malware detection," *IWSPA 2015 - Proc. 2015 ACM Int. Work. Secur. Priv. Anal. Co-located with CODASPY 2015*, pp. 19–26, Mar. 2015, doi: 10.1145/2713579.2713585.

[99] E. Raff and C. Nicholas, "An alternative to NCD for large sequences, lempel-ZiV jaccard distance," *Proc. ACM SIGKDD Int. Conf. Knowl. Discov. Data Min.*, vol. Part F1296, pp. 1007–1015, Aug. 2017, doi: 10.1145/3097983.3098111.

[100] M. Ahmadi, D. Ulyanov, S. Semenov, M. Trofimov, and G. Giacinto, "Novel feature extraction, selection and fusion for effective malware family classification," *CODASPY 2016 - Proc. 6th ACM Conf. Data Appl. Secur. Priv.*, pp. 183–194, Mar. 2016, doi: 10.1145/2857705.2857713.

[101] S. Srakaew, W. Piyanuntcharatsr, S. Adulkasem, and C. Chantrapornchai, "On the Comparison of Malware Detection Methods Using Data Mining with Two Feature Sets," *Int. J. Secur. Its Appl.*, vol. 9, no. 3, pp. 293–318, 2015, doi: 10.14257/ijsia.2015.9.3.23.

[102] A. Shabtai, R. Moskovitch, C. Feher, S. Dolev, and Y. Elovici, "Detecting unknown malicious code by applying classification techniques on OpCode patterns," *Secur. Inform.*, vol. 1, no. 1, pp. 1–22, 2012, doi: 10.1186/2190-8532-1-1.

[103] I. Santos, J. Devesa, F. Brezo, J. Nieves, and P. G. Bringas, "OPEM: A static-dynamic approach for machine-learning-based malware detection," *Adv. Intell. Syst. Comput.*, vol. 189 AISC, pp. 271–280, 2013, doi: 10.1007/978-3-642-33018-6_28.

[104] D. Yuxin and Z. Siyi, "Malware detection based on deep learning algorithm," *Neural Comput. Appl.*, vol. 31, no. 2, pp. 461–472, Jul. 2019, doi: 10.1007/s00521-017-3077-6.

[105] J. Sexton, C. Storlie, and B. Anderson, "Subroutine based detection of APT malware," *J. Comput. Virol. Hacking Tech. 2015 124*, vol. 12, no. 4, pp. 225–233, Dec. 2015, doi: 10.1007/S11416-015-0258-7.

[106] P. Khodamoradi, M. Fazlali, F. Mardukhi, and M. Nosrati, "Heuristic metamorphic malware detection based on statistics of assembly instructions using classification algorithms," *18th CSI Int. Symp. Comput. Archit. Digit. Syst. CADS 2015*, vol. 2015-Janua, pp. 1–6, Jan. 2016, doi: 10.1109/CADS.2015.7377792.

[107] M. Gharacheh, V. Derhami, S. Hashemi, and S. M. H. Fard, "Proposing an HMM-based approach to detect metamorphic malware," *4th Iran. Jt. Congr. Fuzzy Intell. Syst. CFIS 2015*, Jan. 2016, doi: 10.1109/CFIS.2015.7391648.

[108] Dong-Hee Kim, S.-U. Woo, D.-K. Lee, and  and T.-M. Chung, "Static detection of malware and benign executable using machine learning algorithm," no. c, pp. 14–19, 2016.

[109] M. Asquith, "Extremely scalable storage and clustering of malware metadata," *J. Comput. Virol. Hacking Tech.*, vol. 12, no. 2, pp. 49–58, May 2016, doi: 10.1007/S11416-015-0241-3/TABLES/3.

[110] H. Bai, C. Z. Hu, X. C. Jing, N. Li, and X. Y. Wang, "Approach for malware identification using dynamic behaviour and outcome triggering," *IET Inf. Secur.*, vol. 8, no. 2, pp. 140–151, Mar. 2014, doi: 10.1049/IET-IFS.2012.0343.

[111] D. Kirat, L. Nataraj, G. Vigna, and B. S. Manjunath, "SigMal: A static signal processing based malware triage," *ACM Int. Conf. Proceeding Ser.*, pp. 89–98, 2013, doi: 10.1145/2523649.2523682.

[112] G. E. Dahl, J. W. Stokes, L. Deng, and D. Yu, "Large-scale malware classification using random projections and neural networks," *ICASSP, IEEE Int. Conf. Acoust. Speech Signal Process. - Proc.*, pp. 3422–3426, Oct. 2013, doi: 10.1109/ICASSP.2013.6638293.

[113] W. Huang and J. W. Stokes, "MtNet: A Multi-Task Neural Network for Dynamic Malware Classification," *Lect. Notes Comput. Sci. (including Subser. Lect. Notes Artif. Intell. Lect. Notes Bioinformatics)*, vol. 9721, pp. 399–418, 2016, doi: 10.1007/978-3-319-40667-1_20.

[114] R. Islam, R. Tian, L. M. Batten, and S. Versteeg, "Classification of malware based on integrated static and dynamic features," *J. Netw. Comput. Appl.*, vol. 36, no. 2, pp. 646–656, 2013, doi: 10.1016/j.jnca.2012.10.004.

[115] Q. Le, O. Boydell, B. Mac Namee, and M. Scanlon, "Deep learning at the shallow end: Malware classification for non-domain experts," *Digit. Investig.*, vol. 26, pp. S118–S126, Jul. 2018, doi: 10.1016/J.DIIN.2018.04.024.

[116] J. Kinable and O. Kostakis, "Malware classification based on call graph clustering," *J. Comput. Virol.*, vol. 7, no. 4, pp. 233–245, Feb. 2011, doi: 10.1007/s11416-011-0151-y.

[117] M. Hassen and P. K. Chan, "Scalable function call graph-based malware classification," *CODASPY 2017 - Proc. 7th ACM Conf. Data Appl. Secur. Priv.*, pp. 239–248, 2017, doi: 10.1145/3029806.3029824.

[118] M. Eskandari and S. Hashemi, "Metamorphic Malware Detection using Control Flow Graph Mining," *IJCSNS Int. J. Comput. Sci. Netw. Secur.*, vol. 11, no. 12, pp. 1–6, 2011.

[119] P. Faruki, V. Laxmi, M. S. Gaur, and P. Vinod, "Mining control flow graph as API call-grams to detect portable executable malware," *Proc. 5th Int. Conf. Secur. Inf. Networks, SIN'12*, vol. 12, pp. 130–137, 2012, doi: 10.1145/2388576.2388594.

[120] I. Sorokin, "Comparing files using structural entropy," *J. Comput. Virol.*, vol. 7, no. 4, pp. 259–265, Jun. 2011, doi: 10.1007/s11416-

011-0153-9.

[121] D. Baysa, R. M. Low, and M. Stamp, “Structural entropy and metamorphic malware,” *J. Comput. Virol. Hacking Tech. 2013 94*, vol. 9, no. 4, pp. 179–192, Apr. 2013, doi: 10.1007/S11416-013-0185-4.

[122] M. Wojnowicz, G. Chisholm, M. Wolff, and X. Zhao, “Wavelet decomposition of software entropy reveals symptoms of malicious code,” *J. Innov. Digit. Ecosyst.*, vol. 3, no. 2, pp. 130–140, Dec. 2016, doi: 10.1016/j.jides.2016.10.009.

[123] K. Rieck, P. Trinius, C. Willems, and T. Holz, “Automatic analysis of malware behavior using machine learning,” *J. Comput. Secur.*, vol. 19, no. 4, pp. 639–668, Jan. 2011, doi: 10.3233/JCS-2010-0410.

[124] D. Uppal, V. Mehra, and V. Verma, “Basic survey on Malware Analysis, Tools and Techniques,” *Int. J. Comput. Sci. Appl.*, vol. 4, no. 1, pp. 103–112, 2014, doi: 10.5121/ijcsa.2014.4110.

[125] H. S. Galal, Y. B. Mahdy, and M. A. Atiea, “Behavior-based features model for malware detection,” *J. Comput. Virol. Hacking Tech.*, vol. 12, no. 2, pp. 59–67, Jun. 2016, doi: 10.1007/s11416-015-0244-0.

[126] Y. Ki, E. Kim, and H. K. Kim, “A novel approach to detect malware based on API call sequence analysis,” *Int. J. Distrib. Sens. Networks*, vol. 2015, Jun. 2015, doi: 10.1155/2015/659101.

[127] G. Liang, J. Pang, and C. Dai, “A Behavior-Based Malware Variant Classification Technique,” *Int. J. Inf. Educ. Technol.*, vol. 6, no. 4, pp. 291–295, 2016, doi: 10.7763/IJIET.2016.V6.702.

[128] L. Xiaofeng, J. Fangshuo, Z. Xiao, Y. Shengwei, S. Jing, and P. Lio, “ASSCA: API sequence and statistics features combined architecture for malware detection,” *Comput. Networks*, vol. 157, pp. 99–111, Jul. 2019, doi: 10.1016/j.comnet.2019.04.007.

[129] B. Kolosnjaji, A. Zarras, G. Webster, and C. Eckert, “Deep learning for classification of malware system call sequences,” *Lect. Notes Comput. Sci. (including Subser. Lect. Notes Artif. Intell. Lect. Notes Bioinformatics)*, vol. 9992 LNAI, pp. 137–149, Dec. 2016, doi: 10.1007/978-3-319-50127-7_11.

[130] A. A. E. Elhadi, M. A. Maarof, and B. I. A. Barry, “Improving the detection of malware behaviour using simplified data dependent API call graph,” *Int. J. Secur. its Appl.*, vol. 7, no. 5, pp. 29–42, 2013, doi: 10.14257/ijsia.2013.7.5.03.

[131] W. Mao, Z. Cai, D. Towsley, and X. Guan, “Probabilistic Inference on Integrity for Access Behavior Based Malware Detection,” *Lect. Notes Comput. Sci. (including Subser. Lect. Notes Artif. Intell. Lect. Notes Bioinformatics)*, vol. 9404, pp. 155–176, 2015, doi: 10.1007/978-3-319-26362-5_8.

[132] M. Kruczkowski and E. Niewiadomska-Szynkiewicz, “Support vector machine for malware analysis and classification,” *Proc. - 2014 IEEE/WIC/ACM Int. Jt. Conf. Web Intell. Intell. Agent Technol. - Work. WI-IAT 2014*, vol. 2, pp. 415–420, Oct. 2014, doi: 10.1109/WI-IAT.2014.127.

[133] A. Mohaisen, O. Alrawi, and M. Mohaisen, “AMAL: High-fidelity, behavior-based automated malware analysis and classification,” *Comput. Secur.*, vol. 52, pp. 251–266, Jul. 2015, doi: 10.1016/j.cose.2015.04.001.

[134] P. Vadrevu and R. Perdisci, “MAXS: Scaling malware execution with sequential multi-hypothesis testing,” *ASIA CCS 2016 - Proc. 11th ACM Asia Conf. Comput. Commun. Secur.*, pp. 771–782, May 2016, doi: 10.1145/2897845.2897873.

[135] D. Bekerman, B. Shapira, L. Rokach, and A. Bar, “Unknown malware detection using network traffic classification,” *2015 IEEE Conf. Commun. NetworkSecurity, CNS 2015*, pp. 134–142, Dec. 2015, doi: 10.1109/CNS.2015.7346821.

[136] D. Arivudainambi, V. K. Varun, S. C. S., and P. Visu, “Malware traffic classification using principal component analysis and artificial neural network for extreme surveillance,” *Comput. Commun.*, vol. 147, pp. 50–57, Nov. 2019, doi: 10.1016/j.comcom.2019.08.003.

[137] M. Rabbani, Y. L. Wang, R. Khoshkangini, H. Jelodar, R. Zhao, and P. Hu, “A hybrid machine learning approach for malicious behaviour detection and recognition in cloud computing,” *J. Netw. Comput. Appl.*, vol. 151, p. 102507, Feb. 2020, doi: 10.1016/j.jnca.2019.102507.

[138] M. Ghiasi, A. Sami, and Z. Salehi, “Dynamic VSA: a framework for malware detection based on register contents,” *Eng. Appl. Artif. Intell.*, vol. 44, pp. 111–122, Sep. 2015, doi: 10.1016/j.engappai.2015.05.008.

[139] Ç. Yücel and A. Koltuksuz, “Imaging and evaluating the memory access for malware,” *Forensic Sci. Int. Digit. Investig.*, vol. 32, p. 200903, Mar. 2020, doi: 10.1016/J.FSIDI.2019.200903.

[140] X. Liu, Y. Lin, H. Li, and J. Zhang, “A novel method for malware detection on ML-based visualization technique,” *Comput. Secur.*, vol. 89, p. 101682, Feb. 2020, doi: 10.1016/J.COSE.2019.101682.

[141] D. Vasan, M. Alazab, S. Wassan, B. Safaei, and Q. Zheng, “Image-Based malware classification using ensemble of CNN architectures (IMCEC),” *Comput. Secur.*, vol. 92, p. 101748, May 2020, doi: 10.1016/J.COSE.2020.101748.

[142] D. Escudero García and N. DeCastro-García, “Optimal feature configuration for dynamic malware detection,” *Comput. Secur.*, vol. 105, p. 102250, Jun. 2021, doi: 10.1016/j.cose.2021.102250.

[143] C. Storlie, B. Anderson, S. Vander Wiel, D. Quist, C. Hash, and N. Brown, “Stochastic identification of malware with dynamic traces,” *https://doi.org/10.1214/13-AOAS703*, vol. 8, no. 1, pp. 1–18, Mar. 2014, doi: 10.1214/13-AOAS703.

[144] D. Carlin, A. Cowan, P. O’Kane, and S. Sezer, “The Effects of Traditional Anti-Virus Labels on Malware Detection Using Dynamic

Runtime Opcodes," *IEEE Access*, vol. 5, pp. 17742–17752, Sep. 2017, doi: 10.1109/ACCESS.2017.2749538.

[145] M. Alaeiyan, S. Parsa, and M. Conti, "Analysis and classification of context-based malware behavior," *Comput. Commun.*, vol. 136, pp. 76–90, Feb. 2019, doi: 10.1016/j.comcom.2019.01.003.

[146] H. El Merabet and A. Hajraoui, "A survey of malware detection techniques based on machine learning," *Int. J. Adv. Comput. Sci. Appl.*, vol. 10, no. 1, pp. 366–373, 2019, doi: 10.14569/IJACSA.2019.0100148.

[147] F. Mira, A. Brown, and W. Huang, "Novel malware detection methods by using LCS and LCSS," *2016 22nd Int. Conf. Autom. Comput. ICAC 2016 Tackling New Challenges Autom. Comput.*, pp. 554–559, Oct. 2016, doi: 10.1109/ICONAC.2016.7604978.

[148] Z. Markel and M. Bilzor, "Building a machine learning classifier for malware detection," *WATeR 2014 - Proc. 2014 2nd Work. Anti-Malware Test. Res.*, Jan. 2015, doi: 10.1109/WATER.2014.7015757.

[149] K. Shaukat, S. Luo, V. Varadharajan, I. A. Hameed, and M. Xu, "A Survey on Machine Learning Techniques for Cyber Security in the Last Decade," *IEEE Access*, vol. 8, pp. 222310–222354, 2020, doi: 10.1109/ACCESS.2020.3041951.

[150] A. Damodaran, F. Di Troia, C. A. Visaggio, T. H. Austin, and M. Stamp, "A comparison of static, dynamic, and hybrid analysis for malware detection," *J. Comput. Virol. Hacking Tech. 2015 131*, vol. 13, no. 1, pp. 1–12, Dec. 2015, doi: 10.1007/S11416-015-0261-Z.

[151] S. Euh, H. Lee, D. Kim, and D. Hwang, "Comparative analysis of low-dimensional features and tree-based ensembles for malware detection systems," *IEEE Access*, vol. 8, pp. 76796–76808, 2020, doi: 10.1109/ACCESS.2020.2986014.

[152] E. Raff *et al.*, "An investigation of byte n-gram features for malware classification," *J. Comput. Virol. Hacking Tech.*, vol. 14, no. 1, pp. 1–20, Feb. 2018, doi: 10.1007/S11416-016-0283-1/TABLES/14.

[153] Y. Nagano and R. Uda, "Static analysis with paragraph vector for malware detection," *Proc. 11th Int. Conf. Ubiquitous Inf. Manag. Commun. IMCOM 2017*, Jan. 2017, doi: 10.1145/3022227.3022306.

[154] A. Pektaş and T. Acarman, "Classification of malware families based on runtime behaviors," *J. Inf. Secur. Appl.*, vol. 37, pp. 91–100, Dec. 2017, doi: 10.1016/J.JISA.2017.10.005.

[155] A. Namavar Jahromi *et al.*, "An improved two-hidden-layer extreme learning machine for malware hunting," *Comput. Secur.*, vol. 89, 2020, doi: 10.1016/j.cose.2019.101655.

[156] R. Damaševičius, A. Venčkauskas, J. Toldinas, and Š. Grigaliūnas, "Ensemble-based classification using neural networks and machine learning models for windows pe malware detection," *Electron.*, vol. 10, no. 4, pp. 1–26, 2021, doi: 10.3390/electronics10040485.

[157] E. Amer and I. Zelinka, "A dynamic Windows malware detection and prediction method based on contextual understanding of API call sequence," *Comput. Secur.*, vol. 92, p. 101760, May 2020, doi: 10.1016/J.COSE.2020.101760.

[158] P. V. Shijo and A. Salim, "Integrated Static and Dynamic Analysis for Malware Detection," *Procedia Comput. Sci.*, vol. 46, pp. 804–811, Jan. 2015, doi: 10.1016/J.PROCS.2015.02.149.

[159] R. J. Mangialardo and J. C. Duarte, "Integrating Static and Dynamic Malware Analysis Using Machine Learning," *IEEE Lat. Am. Trans.*, vol. 13, no. 9, pp. 3080–3087, Sep. 2015, doi: 10.1109/TLA.2015.7350062.

[160] N. Kumar, S. Mukhopadhyay, M. Gupta, A. Handa, and S. K. Shukla, "Malware classification using early stage behavioral analysis," *Proc. - 2019 14th Asia Jt. Conf. Inf. Secur. AsiaJCIS 2019*, pp. 16–23, Aug. 2019, doi: 10.1109/ASIAJCIS.2019.00-10.

[161] D. Gupta and R. Rani, "Improving malware detection using big data and ensemble learning," *Comput. Electr. Eng.*, vol. 86, p. 106729, Sep. 2020, doi: 10.1016/J.COMPELECENG.2020.106729.

[162] A. Damodaran, F. Di Troia, C. A. Visaggio, T. H. Austin, and M. Stamp, "A comparison of static, dynamic, and hybrid analysis for malware detection," *J. Comput. Virol. Hacking Tech. 2015 131*, vol. 13, no. 1, pp. 1–12, Dec. 2015, doi: 10.1007/S11416-015-0261-Z.

[163] W. Han, J. Xue, Y. Wang, Z. Liu, and Z. Kong, "MalInsight: A systematic profiling based malware detection framework," *J. Netw. Comput. Appl.*, vol. 125, pp. 236–250, Jan. 2019, doi: 10.1016/j.jnca.2018.10.022.